\documentclass[twocolumn,floats,floatfix,amssymb,nofootinbib,prd,superscriptaddress]{revtex4-2}
\usepackage{graphicx,amssymb,amsmath,amsthm,amsfonts,epsfig}
\usepackage[linktocpage,colorlinks=true,citecolor=OliveGreen,linkcolor=Maroon,urlcolor=Maroon]{hyperref}
\usepackage[usenames]{color}
\usepackage{epstopdf}
\usepackage{bm}
\usepackage{dcolumn}
\usepackage[utf8]{inputenc}
\usepackage{latexsym}
\usepackage{rotating}
\usepackage{hyperref}
\usepackage{tabularx}
\usepackage{braket}
\usepackage{color}
\usepackage{esvect}
\usepackage{longtable}
\usepackage{enumerate}
\usepackage{footmisc}
\usepackage{array}
\usepackage{tensor}
\usepackage{mathtools}
\usepackage{url}
\usepackage[dvipsnames]{xcolor}
\usepackage{comment}
\usepackage{multirow}
\usepackage{nicematrix}
\renewcommand{\vec}[1]{\boldsymbol{#1}}

\newcommand{\ben}{\begin{enumerate}}
\newcommand{\een}{\end{enumerate}}

\def\be{\begin{equation}}
\def\ee{\end{equation}}
\def\bea{\begin{eqnarray}}
\def\eea{\end{eqnarray}}
\newcommand{\beq}{\begin{eqnarray}}
\newcommand{\eeq}{\end{eqnarray}} 
\newcommand{\ba}{\begin{align}}
\newcommand{\ea}{\end{align}}
\def\ba{\bar{a}}

\newcommand*{\mline}[1]{%
\begingroup
    \renewcommand*{\arraystretch}{1.5}%
   \begin{tabular}[c]{@{}>{\raggedright\arraybackslash}p{2.5cm}@{}}#1\end{tabular}%
  \endgroup
}

\definecolor{cornellGreen}{HTML}{6EB43F}
\definecolor{cornellRed}{HTML}{B31B1B}
\begin{document}
\title{Impact of a plasma on the relaxation of black holes}
\author{Enrico Cannizzaro}
\affiliation{Dipartimento di Fisica, ``Sapienza" Universit\`{a} di Roma \& Sezione INFN Roma1, Piazzale Aldo Moro 5, 00185, Roma, Italy}
\affiliation{Niels Bohr International Academy, Niels Bohr Institute, Blegdamsvej 17, 2100 Copenhagen, Denmark}
\author{Thomas F.M.~Spieksma}
\affiliation{Niels Bohr International Academy, Niels Bohr Institute, Blegdamsvej 17, 2100 Copenhagen, Denmark}
\author{Vitor Cardoso}
\affiliation{Niels Bohr International Academy, Niels Bohr Institute, Blegdamsvej 17, 2100 Copenhagen, Denmark}
\affiliation{CENTRA, Departamento de F\'{\i}sica, Instituto Superior T\'ecnico -- IST, Universidade de Lisboa -- UL,
Avenida Rovisco Pais 1, 1049 Lisboa, Portugal}
\affiliation{Yukawa Institute for Theoretical Physics, Kyoto University, Kyoto
606-8502, Japan}
\author{Taishi Ikeda}
\affiliation{Niels Bohr International Academy, Niels Bohr Institute, Blegdamsvej 17, 2100 Copenhagen, Denmark}
\begin{abstract}
Our universe is permeated with interstellar plasma, which prevents propagation of low-frequency electromagnetic waves. Here, we show that two dramatic consequences arise out of such suppression;~(i) if plasma permeates the light ring of a black hole, electromagnetic modes are screened entirely from the gravitational-wave signal, changing the black hole spectroscopy paradigm;~(ii) if a near vacuum cavity is formed close to a charged black hole, as expected for near equal-mass mergers, ringdown ``echoes'' are excited. The amplitude of such echoes decays slowly and could thus serve as a silver bullet for plasmas near charged black holes.
\end{abstract}
\maketitle
%
%%%%%%%%%%%%%%%%%%%%%%%%%%%%%%%%%
\noindent {\bf \em Introduction.} 
%%%%%%%%%%%%%%%%%%%%%%%%%%%%%%%%%
The ability to detect gravitational waves (GWs) opened new horizons to advance our understanding of the Universe~\cite{LIGOScientific:2016aoc,LIGOScientific:2018mvr,LIGOScientific:2020ibl,LIGOScientific:2021djp}. GWs probe gravity in the strong-field, dynamical regime \cite{Berti:2015itd,Barack:2018yly,Cardoso:2019rvt,LIGOScientific:2019fpa,LIGOScientific:2020tif,LIGOScientific:2021sio}, they probe environments of compact objects~\cite{Barausse:2014tra,Cardoso:2021wlq,Cardoso:2022whc,Cole:2022yzw,CanevaSantoro:2023aol}, and illuminate the ``dark'' Universe \cite{Bertone:2004pz,Bertone:2018krk,Barack:2018yly,Bertone:2019irm,Cardoso:2019rvt,Brito:2015oca}.
Together with black holes (BHs), GWs hold an exciting potential to search for new interactions or physics. A particularly intriguing possibility concerns charge. Significant amounts of electromagnetic (EM) charge are not expected to survive long for accreting systems (due to selective accretion, Hawking radiation or pair production~\cite{Eardley:1975kp,Gibbons:1975kk,Cardoso:2016olt}), but exceptions exist. A fraction of the primordial BHs produced in the early universe can carry a large amount of charge, suppressing Hawking radiation and potentially allowing for electric or color-charged BHs to survive to our days~\cite{deFreitasPacheco:2023hpb,Alonso-Monsalve:2023brx}. Additionally, BH mergers might be accompanied by strong magnetic fields pushing surrounding plasma to large radii, and preventing neutralisation processes. 
Beyond the realm of Standard Model physics, BHs could be charged in a variety of different models, by circumventing in different ways discharge mechanisms. These models include millicharged dark matter or hidden vector fields, constructed to be viable cold dark matter candidates~\cite{Bai:2019zcd,Kritos:2021nsf,DeRujula:1989fe,Holdom:1985ag,Davidson:2000hf,Dubovsky:2003yn,Sigurdson:2004zp,Gies:2006ca,Gies:2006hv,Burrage:2009yz,Ahlers:2009kh,McDermott:2010pa,Dolgov:2013una,Haas:2014dda,Cardoso:2016olt,Khalil:2018aaj,Caputo:2019tms,Gupta:2021rod,Carullo:2021oxn,Fiorillo:2024upk}. Finally, some BH mimickers are globally neutral while possessing a non-vanishing dipole moment, thus emitting EM radiation. Examples include topological solitons in string-theory fuzzballs scenarios~\cite{Bah:2020ogh, Bah:2021owp}. 

\begin{figure}
    \centering
    \includegraphics[width = 0.95\linewidth]{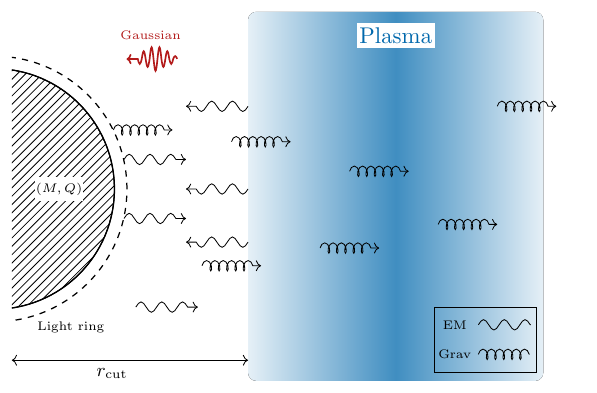}
    \caption{Schematic illustration of our setup:~a charged BH surrounded by plasma is stimulated by external processes (an initial gaussian wavepacket), emitting electromagnetic and gravitational radiation. While GWs are able to travel through the plasma to distant observers, low frequency EM waves are not. Instead they excite further GWs, echoes of the original burst.}    \label{fig:schematicillustration}
\end{figure}
Charge constraints via GW dephasing in the inspiral phase of two compact objects, or via BH spectroscopy assume implicitly that photons propagate freely from source to observer~\cite{Cardoso:2016olt,Khalil:2018aaj,Gupta:2021rod,Carullo:2021oxn}. But the Universe is filled with matter. Even if dilute, the interstellar plasma prevents the propagation of EM waves with frequencies smaller than the plasma frequency, which effectively behaves as an effective mass~\cite{Cardoso:2020nst}:
\begin{equation}
\omega_{\rm p}=\sqrt{\frac{e^2n_{\rm e}}{\epsilon_0 m_{\rm e}}}\sim 1.8\times 10^3\left(\frac{n_{\rm e}}{10^{-3}{\rm cm}^{-3}}\right)^{1/2}\,{\rm rad \,s}^{-1}\,,\label{eq:plasmafreq}
\end{equation}
where $n_{\rm e}$ is the electron number density in the plasma, whereas $m_{\rm e}$ and $e$ are the electron mass and charge, respectively, and $\epsilon_0$ is the vacuum permeability. 

The emission of GWs and EMs
during mergers of compact, charged objects is a coupled phenomenon. Hence, the BH {\it gravitational} spectrum contains EM-driven modes~\cite{Leaver:1990zz,Berti:2009kk}. But if EM modes are unable to propagate, their impact on GW generation and propagation could be important, affecting spectroscopy tests to an unknown degree. 
Motivated by recent progress~\cite{Cardoso:2020nst,Cannizzaro:2020uap,Cannizzaro:2021zbp}, the purpose of this {\it Letter} is to close the gap, by showing from first principles that (i) EM waves are indeed screened by plasma, which filters out EM-led modes from GWs and (ii) in certain plasma-depleted environments, GW echoes are triggered, serving as a clear observational signature of plasmas surrounding charged BHs. A schematic illustration of our setup is shown in Fig.~\ref{fig:schematicillustration}.

We adopt the mostly positive metric signature and use geometrized units in which $G = c = 1$ and Gaussian units for the Maxwell equations. In general, we denote dimensionless quantities with a bar, such as the charge $\overline{Q} = Q/M$ or the plasma frequency $\overline{\omega}_{\rm p}^{\rm (c)} = M \omega_{\rm p}^{\rm (c)}$.

%%%%%%%%%%%%%%%%%%%%%%%%%%
\noindent {\bf \em Setup.} 
%%%%%%%%%%%%%%%%%%%%%%%%%%
We consider a stationary system of charged, gravitating particles, i.e.,~an ``Einstein cluster,'' surrounding a charged BH~\cite{Einstein:1939ms,2012IJMPS..12..146G,Cardoso:2021wlq, Cardoso:2022whc, Feng:2022evy}. The key idea behind the model is to perform an angular average over particles in circular motion in all possible orientations, which is equivalent to considering an anisotropic, static fluid with a non-vanishing tangential pressure. Focusing on a fluid consisting of electrons, the stress-energy tensor reads\footnote{The following discussion also applies to millicharged dark matter. For clarity purposes, we focus on electrons.}
\begin{equation}
\label{eq:plasmaSET}
    T^{\rm p}_{\mu\nu}=(\rho+P_{\rm t}) v_\mu v_\nu+P_{\rm t}(g_{\mu\nu}-r_\mu r_\nu)\,,
\end{equation}
where $\rho=n_{\rm e} m_{\rm e}$ is the energy density of the fluid, $v^\mu$ the four-velocity, $P_{\rm t}$ the tangential pressure, $g_{\mu \nu}$ the metric of the underlying spacetime and $r^\mu$ a unit vector in the radial direction. 

We then consider Einstein-Maxwell theory in the presence of this fluid. The relevant field equations are
\begin{equation}
\label{eq:einstein_maxwell}
G_{\mu\nu}= 8 \pi \left(T_{\mu\nu}^{\rm EM}+T_{\mu\nu}^{\rm p}\right)\,, \quad \nabla_\nu F^{\mu\nu}=j^\mu\,,
\end{equation}
where $G_{\mu\nu}$ and $F_{\mu\nu}$ are the Einstein and Maxwell tensor, respectively, $j^\mu=e n_{\rm e} v^\mu$ the plasma current and $T_{\mu\nu}^{\rm EM}$ is the stress-energy tensor for the EM sector:
\begin{equation}
\begin{aligned}
T_{\mu\nu}^{\rm EM}&=\frac{1}{4 \pi}\left(g^{\rho \sigma}F_{\rho \mu}F_{\sigma \nu}-\frac{1}{4}g_{\mu\nu}F_{\rho \sigma}F^{\rho \sigma}\right)\,.
\end{aligned}
\end{equation}
Finally, to close the system, the momentum and continuity equation of the charged fluid are needed, which are derived from the conservation of the stress-energy tensors and the current,
\begin{equation}
\label{eq:momentum_continuity}
\nabla^\nu T^{\rm p}_{\mu\nu}=e n_{\rm e} F_{\mu\nu}v^\nu\,, \quad \nabla_\mu (n_{\rm e} v^\mu)=0\,.
\end{equation}
In the following, we ignore backreaction of the fluid in the Einstein and Maxwell equations as source terms are suppressed by the large charge-to-mass ratio of the electron, and energy densities of astrophysical fluids are small. In addition, astrophysical plasmas typically include ions, which induce a current with the opposite sign in the Maxwell equations and can be considered a stationary, neutralizing background~\cite{Cannizzaro:2020uap,Cannizzaro:2021zbp,Cannizzaro:2023ltu,Spieksma:2023vwl}. Accordingly, one obtains the Reissner-Nordstr\"{o}m (RN) solution from Eqs.~\eqref{eq:einstein_maxwell}, describing a spherically symmetric, charged BH:
\begin{equation}
\label{eq:RN_Sol}
\begin{aligned}
\mathrm{d}s^2 & =-f\mathrm{d} t^2+f^{-1} \mathrm{d} r^2+r^2\mathrm{d}\Omega^{2}\,, \\
&\text{with}\quad f= 1-\frac{2 M}{r}+\frac{Q^2}{r^2}\,, 
\end{aligned}
\end{equation}
where $M$ and $Q$ are BH mass and charge, respectively, and $\mathrm{d}\Omega^{2}$ is the metric on the 2-sphere. The event horizon is at $r_+=M+\sqrt{M^2-Q^2}$ and the light ring at $r_{\rm LR}=3M/2+\sqrt{9M^{2}-8Q^{2}}/2$. We will assume a non-relativistic fluid, i.e., $P_{\rm t}\ll\rho$, such that~\eqref{eq:plasmaSET} reduces to $T^{\rm p}_{\mu\nu}\approx \rho v_\mu v_\nu+P_{\rm t}(g_{\mu\nu}-r_\mu r_\nu)$, and the left hand side of the momentum equation~\eqref{eq:momentum_continuity} resembles the non-relativistic Euler equation. Solving the momentum equation~\eqref{eq:momentum_continuity} then yields the tangential pressure
\begin{equation}
\label{eq:tangentialpressure}
    P_{\rm t}=-\frac{n_{\rm e}(e Q r\sqrt{f}+(Q^2-Mr)m_{\rm e})}{Q^2-3Mr+2r^2}\,.
\end{equation}
For the non-relativistic assumption $P_{\rm t} \ll \rho = n_{\rm e} m_{\rm e}$ to hold, we must have either $Q/M<m_{\rm e}/e$ or $r\gg M$. Given the large charge-to-mass ratio of electrons ($e/m_{\rm e}\approx 10^{22}$), the former condition is only satisfied for extremely weakly charged BHs. Nevertheless, a number of effects can affect this outcome, such as magnetic fields, the formation of a cavity in the  plasma due to mergers~\cite{Armitage:2005xq,Artymowicz:1994bw,Armitage:2005xq,Grobner:2020drr,Farris:2014zjo,Ishibashi:2020zzy,2013MNRAS.436.2997D,2017A&A...598A..43C} or the partial screening of the BH charge by plasma over a Debye length~\cite{Feng:2022evy, Alonso-Monsalve:2023jfq}. In the following, we consider high values of $Q$ as a proxy to model these scenarios, which are too complicated to be included in a self-consistent way. Moreover, for millicharged dark matter, the charge-to-mass ratio of the particles can be arbitrarily small. 

Considering the full momentum equation allows us to study the relativistic regime as well. As detailed in the Supplemental Material, this regime generates similar results, albeit with a largely suppressed effective mass. Interestingly, this suppression can be understood as a form of strong-field {\it transparency} for relativistic plasmas, induced by the background charge $Q$~\cite{KawDawson1970, Cardoso:2020nst}. Upon tuning the plasma density, we thus expect the same phenomenology to hold. Furthermore, at large distances the transparency effect vanishes, yielding the standard effective mass.

Consider now the linearization of the field equations~\eqref{eq:einstein_maxwell} around the RN geometry, the background fields and fluid variables. Perturbations can then be decomposed in two sectors---{\it axial} (or odd) and {\it polar} (or even)---depending on their behaviour under parity transformations. These two sectors decouple in spherically symmetric geometries (see Supplemental Material)~\cite{Regge:1957td,Zerilli:1970wzz,Zerilli:1970se,Zerilli:1974ai, Pani:2013wsa, Rosa:2011my, Baryakhtar:2017ngi}.

The axial sector is completely determined by two functions, a Moncrief-like ``master gravitational'' variable $\Psi$~\cite{Moncrief:1974ng,PhysRevD.9.2707,Moncrief:1975sb} and a ``master EM'' variable $u_4$, which obey a coupled set of second order, partial differential wavelike equations,
\begin{equation}
\label{eq:wavelike-eqn}
\begin{aligned}
\hat{\mathcal{L}}\Psi &=\Bigg(\frac{4 Q^4}{r^6}+ \frac{Q^2(-14 M + r (4+\lambda))}{r^5} \\&+ \left(1-\frac{2M}{r}\right)\left[\frac{\lambda}{r^{2}}-\frac{6M}{r^{3}}
\right]\Bigg)\Psi-\frac{8 Q f}{r^3 \lambda} u_{4}\,, \\
\hat{\mathcal{L}}u_{4}&=f\left(\omega_{\rm p}^2 + \frac{\lambda}{r^{2}} + \frac{4 Q^2}{r^4}\right) u_{4}-\frac{(\ell-1)\lambda(\ell+2) Q f}{2 r^3}\Psi\,,
\end{aligned}
\end{equation}
where $\lambda=\ell(\ell+1)$, $\hat{\mathcal{L}} = \partial^{2}/\partial r_*^{2} - \partial^{2}/\partial t^{2}$ and the tortoise coordinate is defined as $\mathrm{d}r_*/\mathrm{d}r = f^{-1}$. Note that in the limit $Q\rightarrow 0$, the equations decouple:~the first one reduces to the Regge-Wheeler equation while the second one coincides with the axial mode of an EM field in Schwarzschild in the presence of plasma~\cite{Cannizzaro:2020uap}. 

The polar sector is more intricate, with EM and fluid perturbations being coupled. As detailed in the Supplemental Material, at large radii and neglecting metric fluctuations, we recover the dispersion relation $(\omega^2-k^2-\omega_{\rm p}^2)\,\delta F_{12}=0$, where $k$ is the wave vector in Fourier space and $\delta F_{12}$ the perturbed Maxwell tensor. The plasma frequency thus acts as an effective mass for the propagating degree of freedom in the polar sector. As the dynamics emerging in the axial sector are precisely contingent upon this fact, we expect the phenomenology to be similar~\cite{Cannizzaro:2020uap, Cannizzaro:2021zbp} and we hereafter focus only on the axial sector.

%%%%%%%%%%%%%%%%%%%%%%%%%%%%%%%%%%%%
\noindent {\bf \em Initial conditions, plasma profile and numerical procedure.}
%%%%%%%%%%%%%%%%%%%%%%%%%%%%%%%%%%%%
We evolve the wavelike equations \eqref{eq:wavelike-eqn} in time with a two-step Lax-Wendroff algorithm that uses second-order finite differences~\cite{Zenginoglu:2011zz}, following earlier work~\cite{Krivan_1997,Pazos_valos_2005,Zenginoglu:2011zz,Zenginoglu:2012us, Cardoso:2021vjq}. Our grid is uniformly spaced in tortoise coordinates $r_*$, with the boundaries placed sufficiently far away such that boundary effects cannot have an impact on the evolution of the system at the extraction radius. Our code shows second-order convergence (see Supplemental Material).

We consider a plasma profile truncated at a radius $r_{\rm cut}$, smoothened by a sigmoid-like function:
\begin{equation}
\label{eq:plasma-profile2}
\omega_{\rm p} = \omega^{\rm (c)}_{\rm p} \frac{1}{1+e^{- (r-r_{\rm cut})/d}}\,.
\end{equation}
Here, $\omega^{\rm (c)}_{\rm p}$ is the (constant) amplitude of the plasma barrier and $d$ determines how ``sharp'' the cut is. We choose $d = M$, but we verified that the results are not sensitive to this parameter.\footnote{As we will see, the outcome depends on a critical value for $\omega_{\rm p}$ (the fundamental EM QNM), making the density distribution after the barrier ($r > r_{\rm cut}$) or a tenuous plasma before the barrier ($r < r_{\rm cut}$) unimportant for the phenomenology.} Profile \eqref{eq:plasma-profile2} allows us to consider two distinct scenarios;~(i) plasmas that ``permeate'' the light ring $(r_{\rm cut} < r_{\scalebox{0.60}{$\mathrm{LR}$}})$, hence possibly affecting the {\it generation} of quasi-normal modes (QNMs) and (ii)~plasmas localized away from the BH $(r_{\rm cut} \gg r_{\scalebox{0.60}{$\mathrm{LR}$}})$, affecting at most the {\it propagation} of the signal. We consider the initial conditions $\Psi(0,r) =\Psi_0,u_{4}(0,r)=u_{40}$ with~\cite{Cardoso:2020nst}
\begin{equation}
\label{eq:ID}
\begin{aligned}
\mkern-22mu(\Psi_0,u_{40}) &= (A_{\rm g},A_{\rm EM})\exp{\left[-\frac{(r_*-r_0)^2}{2 \sigma^2}-i\Omega_{0} r_*\right]} \,, \\
\partial_t \Psi_0 &= -i\Omega_0 \Psi_0\,, \quad \partial_t u_{40} = -i\Omega_0 u_{40}\,,
\end{aligned}
\end{equation}
where $(A_{\rm g}, A_{\rm EM})=(1,0), (0,1), (1,1)$ for $\mathrm{ID}_{\rm g},\mathrm{ID}_{\rm EM}$ and $\mathrm{ID}_{\rm 2}$, respectively. Throughout this work, we initialize at $r_0 = 20M$ and we extract the signal at $r_{\rm ext} = 300M$. We pick $\sigma = 4.0M$ and wavepacket frequency $\Omega_{0} = 0.1$, yet tested extensively that our results are independent of these factors.

%%%%%%%%%%%%%%%%%%%%%%%%%%%%%%%%%%%%%%%%%%%%%
\noindent {\bf \em Impact of plasma on QNMs.}
%%%%%%%%%%%%%%%%%%%%%%%%%%%%%%%%%%%%%%%%%%%%%
%
\begin{figure}
    \centering
    \includegraphics[width = 0.95\linewidth]{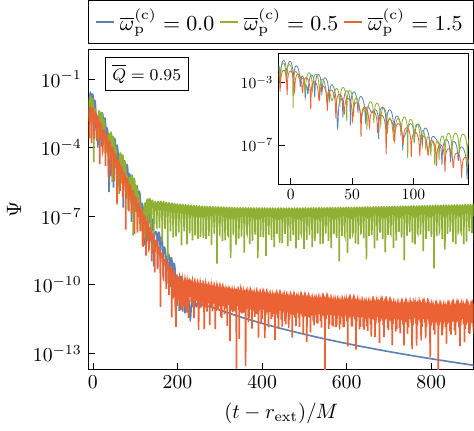}
    \caption{GWs for $\mathrm{ID}_{\rm EM}$ and $\overline{Q} = 0.95$. For sufficiently dense plasmas (when $\overline{\omega}^{\rm (c)}_{\rm p}$ exceeds the EM QNM frequency), the EM mode is screened. This is apparent in the inset:~only one---gravitational-led mode---is present at large plasma frequency. Associated QNM frequencies can be found in Table~\ref{tab:frequencies}. At late times, long-lived modes are excited, in contrast to the usual power-law tail in vacuum. These originate from the quasi-bound states formed in the EM sector and are subsequently imprinted onto the GW signal.} \label{fig:contaminationID_EM}
\end{figure}
\begin{table}[t]
\centering
\renewcommand*{\arraystretch}{1.5}
\resizebox{0.98\linewidth}{!}{
\tabcolsep=0.17cm
\begin{tabular}{|c||c|c|} 
\hline
$\overline{Q} = 0.95$ & $\overline{\omega}_{\rm QNM}$ (time-domain) & $\overline{\omega}_{\rm QNM}$ (frequency-domain) \\
\hline \hline
$\overline{\omega}^{\rm (c)}_{\rm p} = 0.0$ 
& \mline{\begin{tabular}[t]{@{}c@{}}$0.42170~(0.086647)$\\$0.65475~(0.094609)$\end{tabular}} & \mline{\begin{tabular}[t]{@{}c@{}}$0.42169~(0.086659)$\\$0.65476~(0.094605)$\end{tabular}} \\ 
\hline
$\overline{\omega}^{\rm (c)}_{\rm p} = 1.5$  &\mline{$0.45902~(0.090143)$} & \mline{$0.45902~(0.090146)$} \\
\hline
\end{tabular}}
\caption{Real (imaginary) part of the fundamental QNM frequencies as calculated from a time- and frequency-domain approach. We consider: (i) no plasma ({\it top row}), and we find two modes contributing to the signal and (ii) plasma ({\it bottom row}), for which the EM mode is screened and only the gravitational one remains, with a shifted frequency. We obtain similar results in the time-domain, regardless of the chosen $\mathrm{ID}$.}
\label{tab:frequencies}
\end{table}
When plasma permeates the light ring, BH relaxation is expected to change {\it in the EM channel}. We indeed find a total suppression of the EM signal at large distances for large $\omega_{\rm p}$. However, we find something more significant, summarized in Fig.~\ref{fig:contaminationID_EM}, which shows the {\it gravitational} waveform for $\overline{Q}=0.95$ and different plasma frequencies $\omega^{\rm (c)}_{\rm p}$. 
In absence of plasma, the signal is described by a superposition of gravitational- and EM-led modes, clearly visible (see inset) due to the high coupling $Q$. A best-fit to the signal shows the presence of two dominant modes, with (complex) frequencies reported in Table~\ref{tab:frequencies}. 
The plasma suppresses propagation of EM modes when $\omega_{\rm p}$ exceeds the fundamental EM QNM frequency. Our results show that the coupling to GWs also affects the gravitational signal to an important degree. In fact, as apparent in Fig.~\ref{fig:contaminationID_EM}, GWs now carry mostly a single gravitational-led mode (red line), but with a shifted frequency, see Table~\ref{tab:frequencies}. This shift is surprising, and it originates from the coupling between gravity and electromagnetism. The presence of plasma thus affects the QNM frequencies of the gravitational signal.    

We confirm these results by frequency-domain calculations (where QNMs are obtained by direct integration with a shooting method) in Table \ref{tab:frequencies}. Note that we impose purely outgoing boundary conditions at infinity in vacuum, while in the presence of plasma, we consider exponentially decaying EM modes at large distances, to account for quasi-bound states (QBS). Clearly, the results from time- and frequency-domain are in good agreement.

On longer timescales, EM QBSs are formed in the presence of plasma (see Supplemental Material)~\cite{Lingetti:2022psy,Dima:2020rzg}, which ``pollute'' the gravitational signal. These are long-lived states which are prevented from leaking to infinity due to the plasma effective mass, and are thus similar to QBS of massive fundamental fields~\cite{Brito:2015oca}. At late times, we indeed observe a signal ringing at a frequency comparable (yet slightly smaller) than the plasma frequency $\omega_{\rm R} \lesssim \omega_{\rm p}$.
As the plasma frequency is increased, the QBSs form at progressively late times, and thus at lower amplitudes, unreachable for observations. This phenomenology is similar to the toy model considered in Ref.~\cite{Cardoso:2020nst}, but here explored from first principles.

%%%%%%%%%%%%%%%%%%%%%%%%%%%%%%%%%%%%%%%%%%%%%%%%%%%%%
\noindent {\bf \em Propagation:~echoes in waveforms.} 
%%%%%%%%%%%%%%%%%%%%%%%%%%%%%%%%%%%%%%%%%%%%%%%%%%%%%
%
\begin{figure}[t]
    \centering
    \includegraphics[width = 0.95\linewidth]{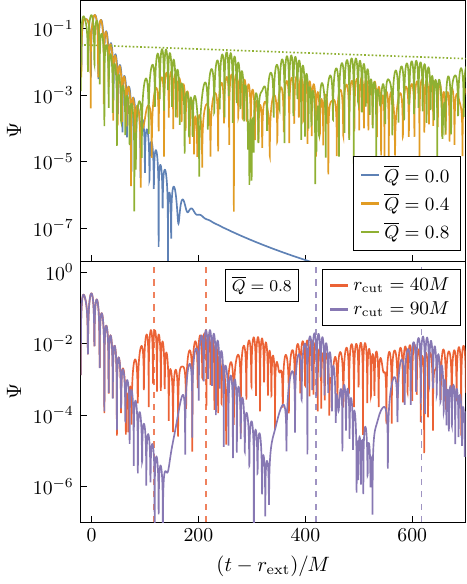}
    \caption{GW signal generated in the presence of a plasma localized away from the BH. In the {\it top panel} $r_{\rm cut} = 40M$; the amplitude of the echoes increases concurrently with the coupling $Q$. Dotted line indicates the decay rate of the signal, $M\Gamma = -0.00133$, as predicted from \eqref{eq:amplitude-fall}. {\it Bottom panel} illustrates how the echo timescale depends on the position of the plasma barrier. The estimates from \eqref{eqn:timescale-barrier} are indicated by vertical dashed lines. Both panels are initialized with $\mathrm{\mathrm{ID}_{\rm g}}$ and $\overline{\omega}^{\rm (c)}_{\rm p} = 1.5$.}
    \label{fig:echoes}
\end{figure}
When the plasma is localized away from the BH, new phenomenology emerges. BH ringdown is associated mostly with light ring physics, hence prompt ringdown is no longer affected~\cite{Cardoso:2016rao,Cardoso:2019rvt}. However, upon exciting the BH, both EM and GWs travel outwards. While GWs travel through the plasma, EM waves are reflected, interacting with the BH again and exciting one more stage of ringdown and corresponding GW ``echoes''. Such echoes have been found before in the context of (near-) horizon quantum structures~\cite{Cardoso:2016rao,Cardoso:2016oxy,Oshita:2018fqu,Wang:2019rcf}, exotic states of matter in ultracompact/neutron stars~\cite{Ferrari:2000sr,Pani:2018flj,Buoninfante:2019swn} or modified theories of gravity~\cite{Buoninfante:2019teo,Delhom:2019btt,Zhang:2017jze} (see~\cite{Cardoso:2017cqb,Cardoso:2019rvt} for reviews). We find them in a General Relativity setting.

In the top panel of Fig.~\ref{fig:echoes}, we show the GW signal for different BH charge. In contrast to vacuum (where exponential ringdown gives way to a power-law tail), in the presence of plasma prompt ringdown is followed by echoes of the original burst. For higher BH charge, the reflected EM signal is more strongly coupled, increasing the amplitude of the GW echoes. 

The main features of the echoing signal are simple to understand. The time between consecutive echoes can be estimated as (for $r_{\rm cut} \gg r_{\scalebox{0.60}{$\mathrm{LR}$}}$)
\begin{equation}
\label{eqn:timescale-barrier}
\Delta t = 2\int_{r_{\scalebox{0.60}{$\mathrm{LR}$}}}^{r_{\rm cut}} \frac{\mathrm{d}r}{f(r)} \approx 2r_{\rm cut}\,,
\end{equation}
This interval is shown by the vertical dashed lines in Fig.~\ref{fig:echoes} and clearly in good agreement with the numerics.

The echo amplitude decays in time, since the BH absorbs part of the reflected waves, and part of the energy is carried to infinity by GWs. The amplitude $A(t)$ of trapped modes in a cavity of length $\sim r_{\rm cut}$ is expected to fall off as  
\begin{equation}
\label{eq:amplitude-fall}
A(t) = A_0 e^{-\Gamma t}\,, \quad \text{with} \quad \Gamma \sim \frac{{\cal A}^2}{r_{\rm cut}}\,,
\end{equation}
where $A_0$ is the initial amplitude and ${\cal A}^2$ the absorption coefficient of the BH (neglecting losses to GWs at infinity). For simplicity, we take the absorption coefficient of low-frequency monochromatic waves for neutral BHs, given by ${\cal A}^2\sim 256(M\omega_{\rm R})^6/225$~\cite{Starobinskil:1974nkd,Brito:2015oca}, where $\omega_{\rm R}$ is the frequency of the trapped EM waves. As the BH absorbs high-frequency modes first, the decay rate decreases over time, asymptoting to a QBS, while the trapped wavepacket broadens. Taking $\omega_{\rm R}$ as the highest-frequency peak in the spectrum, we obtain a decay rate~\eqref{eq:amplitude-fall} in agreement of $\mathcal{O}(1)$ for the first few echoes in Fig.~\ref{fig:echoes}. We confirmed that at later times, the high frequency components of the EM field are indeed lost and the decay rate is decreased accordingly.
A similar phenomenology can be found for any mechanism that places a BHs in a confining box, e.g.~AdS BHs where the AdS radius is much larger than the horizon radius, or Ernst BHs immersed in a magnetic field $B \ll 1/M$~\cite{Brito:2014nja}.

%%%%%%%%%%%%%%%%%%%%%%%%%%%%%%%%%
\noindent {\bf \em  Conclusions.}
%%%%%%%%%%%%%%%%%%%%%%%%%%%%%%%%%
Plasmas are ubiquitous in the universe, but their impact on our ability to do precision GW physics is poorly understood. We have studied plasma physics in curved spacetime from first principles, capturing their impact on the ringdown of charged BHs. Our results are surprising at first sight. We find an important impact of plasma physics on the {\it gravitational waves} generated by charged BHs, changing BH spectroscopy to a measurable extend. We see a ringing frequency going up, and the lifetime of the ringdown going down, a behavior that would be important to dissect.
We also find that plasmas may trigger measurable echoes in GWs. As the amplitude of these echoes decays slowly, they could be in reach of current or future detectors. 

In our work, we focused on values of the plasma frequency $\omega_{\rm p}\sim \mathcal{O}(1/M)$. Note however, that larger values yield similar outcomes. Specifically, a denser plasma present at the light ring causes a greater shift in the gravitational QNM frequency, while a denser plasma localized outside the light ring increases the height of the plasma barrier, making the reflection of photons and thus the echoes, more prominent. Most of our results would also apply to magnetic BHs, which share many similarities with charged BHs in the ringdown phase~\cite{Pereniguez:2023wxf,Dyson:2023ujk}. Finally, as a byproduct, our work introduced a complete framework to describe the behaviour of plasmas around charged compact objects in the (non)-relativistic regime. To simplify our analysis, we modelled the background plasma as a non-relativistic fluid. A complete approach would require consistently evolving the background plasma motion, which is challenging given the large charge-to-mass ratio of electrons (and ions).

\noindent {\bf \em Acknowledgments.} 
We thank Paolo Pani for feedback on the manuscript as well as Gregorio Carullo, Marina De Amicis and David Pere\~{n}iguez for useful conversations. 
We acknowledge support by VILLUM Foundation (grant no. VIL37766) and the DNRF Chair program (grant no. DNRF162) by the Danish National Research Foundation.
V.C.\ is a Villum Investigator and a DNRF Chair.  
V.C. acknowledges financial support provided under the European Union’s H2020 ERC Advanced Grant “Black holes: gravitational engines of discovery” grant agreement no. Gravitas–101052587. 
Views and opinions expressed are however those of the author only and do not necessarily reflect those of the European Union or the European Research Council. Neither the European Union nor the granting authority can be held responsible for them.
This project has received funding from the European Union's Horizon 2020 research and innovation programme under the Marie Sklodowska-Curie grant agreement No 101007855 and No 101131233.
\bibliography{References}

\begin{thebibliography}{96}%
\makeatletter
\providecommand \@ifxundefined [1]{%
 \@ifx{#1\undefined}
}%
\providecommand \@ifnum [1]{%
 \ifnum #1\expandafter \@firstoftwo
 \else \expandafter \@secondoftwo
 \fi
}%
\providecommand \@ifx [1]{%
 \ifx #1\expandafter \@firstoftwo
 \else \expandafter \@secondoftwo
 \fi
}%
\providecommand \natexlab [1]{#1}%
\providecommand \enquote  [1]{``#1''}%
\providecommand \bibnamefont  [1]{#1}%
\providecommand \bibfnamefont [1]{#1}%
\providecommand \citenamefont [1]{#1}%
\providecommand \href@noop [0]{\@secondoftwo}%
\providecommand \href [0]{\begingroup \@sanitize@url \@href}%
\providecommand \@href[1]{\@@startlink{#1}\@@href}%
\providecommand \@@href[1]{\endgroup#1\@@endlink}%
\providecommand \@sanitize@url [0]{\catcode `\\12\catcode `\$12\catcode
  `\&12\catcode `\#12\catcode `\^12\catcode `\_12\catcode `\%12\relax}%
\providecommand \@@startlink[1]{}%
\providecommand \@@endlink[0]{}%
\providecommand \url  [0]{\begingroup\@sanitize@url \@url }%
\providecommand \@url [1]{\endgroup\@href {#1}{\urlprefix }}%
\providecommand \urlprefix  [0]{URL }%
\providecommand \Eprint [0]{\href }%
\providecommand \doibase [0]{http://dx.doi.org/}%
\providecommand \selectlanguage [0]{\@gobble}%
\providecommand \bibinfo  [0]{\@secondoftwo}%
\providecommand \bibfield  [0]{\@secondoftwo}%
\providecommand \translation [1]{[#1]}%
\providecommand \BibitemOpen [0]{}%
\providecommand \bibitemStop [0]{}%
\providecommand \bibitemNoStop [0]{.\EOS\space}%
\providecommand \EOS [0]{\spacefactor3000\relax}%
\providecommand \BibitemShut  [1]{\csname bibitem#1\endcsname}%
\let\auto@bib@innerbib\@empty
%</preamble>
\bibitem [{\citenamefont {Abbott}\ \emph {et~al.}(2016)\citenamefont {Abbott}
  \emph {et~al.}}]{LIGOScientific:2016aoc}%
  \BibitemOpen
  \bibfield  {author} {\bibinfo {author} {\bibfnamefont {B.~P.}\ \bibnamefont
  {Abbott}} \emph {et~al.} (\bibinfo {collaboration} {LIGO Scientific,
  Virgo}),\ }\href {\doibase 10.1103/PhysRevLett.116.061102} {\bibfield
  {journal} {\bibinfo  {journal} {Phys. Rev. Lett.}\ }\textbf {\bibinfo
  {volume} {116}},\ \bibinfo {pages} {061102} (\bibinfo {year} {2016})},\
  \Eprint {http://arxiv.org/abs/1602.03837} {arXiv:1602.03837 [gr-qc]}
  \BibitemShut {NoStop}%
\bibitem [{\citenamefont {Abbott}\ \emph
  {et~al.}(2019{\natexlab{a}})\citenamefont {Abbott} \emph
  {et~al.}}]{LIGOScientific:2018mvr}%
  \BibitemOpen
  \bibfield  {author} {\bibinfo {author} {\bibfnamefont {B.~P.}\ \bibnamefont
  {Abbott}} \emph {et~al.} (\bibinfo {collaboration} {LIGO Scientific,
  Virgo}),\ }\href {\doibase 10.1103/PhysRevX.9.031040} {\bibfield  {journal}
  {\bibinfo  {journal} {Phys. Rev. X}\ }\textbf {\bibinfo {volume} {9}},\
  \bibinfo {pages} {031040} (\bibinfo {year} {2019}{\natexlab{a}})},\ \Eprint
  {http://arxiv.org/abs/1811.12907} {arXiv:1811.12907 [astro-ph.HE]}
  \BibitemShut {NoStop}%
\bibitem [{\citenamefont {Abbott}\ \emph
  {et~al.}(2021{\natexlab{a}})\citenamefont {Abbott} \emph
  {et~al.}}]{LIGOScientific:2020ibl}%
  \BibitemOpen
  \bibfield  {author} {\bibinfo {author} {\bibfnamefont {R.}~\bibnamefont
  {Abbott}} \emph {et~al.} (\bibinfo {collaboration} {LIGO Scientific,
  Virgo}),\ }\href {\doibase 10.1103/PhysRevX.11.021053} {\bibfield  {journal}
  {\bibinfo  {journal} {Phys. Rev. X}\ }\textbf {\bibinfo {volume} {11}},\
  \bibinfo {pages} {021053} (\bibinfo {year} {2021}{\natexlab{a}})},\ \Eprint
  {http://arxiv.org/abs/2010.14527} {arXiv:2010.14527 [gr-qc]} \BibitemShut
  {NoStop}%
\bibitem [{\citenamefont {Abbott}\ \emph
  {et~al.}(2021{\natexlab{b}})\citenamefont {Abbott} \emph
  {et~al.}}]{LIGOScientific:2021djp}%
  \BibitemOpen
  \bibfield  {author} {\bibinfo {author} {\bibfnamefont {R.}~\bibnamefont
  {Abbott}} \emph {et~al.} (\bibinfo {collaboration} {LIGO Scientific, VIRGO,
  KAGRA}),\ }\href@noop {} {\  (\bibinfo {year} {2021}{\natexlab{b}})},\
  \Eprint {http://arxiv.org/abs/2111.03606} {arXiv:2111.03606 [gr-qc]}
  \BibitemShut {NoStop}%
\bibitem [{\citenamefont {Berti}\ \emph {et~al.}(2015)\citenamefont {Berti}
  \emph {et~al.}}]{Berti:2015itd}%
  \BibitemOpen
  \bibfield  {author} {\bibinfo {author} {\bibfnamefont {E.}~\bibnamefont
  {Berti}} \emph {et~al.},\ }\href {\doibase 10.1088/0264-9381/32/24/243001}
  {\bibfield  {journal} {\bibinfo  {journal} {Class. Quant. Grav.}\ }\textbf
  {\bibinfo {volume} {32}},\ \bibinfo {pages} {243001} (\bibinfo {year}
  {2015})},\ \Eprint {http://arxiv.org/abs/1501.07274} {arXiv:1501.07274
  [gr-qc]} \BibitemShut {NoStop}%
%%CITATION = ARXIV:1501.07274;%%
\bibitem [{\citenamefont {Barack}\ \emph {et~al.}(2019)\citenamefont {Barack}
  \emph {et~al.}}]{Barack:2018yly}%
  \BibitemOpen
  \bibfield  {author} {\bibinfo {author} {\bibfnamefont {L.}~\bibnamefont
  {Barack}} \emph {et~al.},\ }\href {\doibase 10.1088/1361-6382/ab0587}
  {\bibfield  {journal} {\bibinfo  {journal} {Class. Quant. Grav.}\ }\textbf
  {\bibinfo {volume} {36}},\ \bibinfo {pages} {143001} (\bibinfo {year}
  {2019})},\ \Eprint {http://arxiv.org/abs/1806.05195} {arXiv:1806.05195
  [gr-qc]} \BibitemShut {NoStop}%
\bibitem [{\citenamefont {Cardoso}\ and\ \citenamefont
  {Pani}(2019)}]{Cardoso:2019rvt}%
  \BibitemOpen
  \bibfield  {author} {\bibinfo {author} {\bibfnamefont {V.}~\bibnamefont
  {Cardoso}}\ and\ \bibinfo {author} {\bibfnamefont {P.}~\bibnamefont {Pani}},\
  }\href {\doibase 10.1007/s41114-019-0020-4} {\bibfield  {journal} {\bibinfo
  {journal} {Living Rev. Rel.}\ }\textbf {\bibinfo {volume} {22}},\ \bibinfo
  {pages} {4} (\bibinfo {year} {2019})},\ \Eprint
  {http://arxiv.org/abs/1904.05363} {arXiv:1904.05363 [gr-qc]} \BibitemShut
  {NoStop}%
\bibitem [{\citenamefont {Abbott}\ \emph
  {et~al.}(2019{\natexlab{b}})\citenamefont {Abbott} \emph
  {et~al.}}]{LIGOScientific:2019fpa}%
  \BibitemOpen
  \bibfield  {author} {\bibinfo {author} {\bibfnamefont {B.~P.}\ \bibnamefont
  {Abbott}} \emph {et~al.} (\bibinfo {collaboration} {LIGO Scientific,
  Virgo}),\ }\href {\doibase 10.1103/PhysRevD.100.104036} {\bibfield  {journal}
  {\bibinfo  {journal} {Phys. Rev. D}\ }\textbf {\bibinfo {volume} {100}},\
  \bibinfo {pages} {104036} (\bibinfo {year} {2019}{\natexlab{b}})},\ \Eprint
  {http://arxiv.org/abs/1903.04467} {arXiv:1903.04467 [gr-qc]} \BibitemShut
  {NoStop}%
\bibitem [{\citenamefont {Abbott}\ \emph
  {et~al.}(2021{\natexlab{c}})\citenamefont {Abbott} \emph
  {et~al.}}]{LIGOScientific:2020tif}%
  \BibitemOpen
  \bibfield  {author} {\bibinfo {author} {\bibfnamefont {R.}~\bibnamefont
  {Abbott}} \emph {et~al.} (\bibinfo {collaboration} {LIGO Scientific,
  Virgo}),\ }\href {\doibase 10.1103/PhysRevD.103.122002} {\bibfield  {journal}
  {\bibinfo  {journal} {Phys. Rev. D}\ }\textbf {\bibinfo {volume} {103}},\
  \bibinfo {pages} {122002} (\bibinfo {year} {2021}{\natexlab{c}})},\ \Eprint
  {http://arxiv.org/abs/2010.14529} {arXiv:2010.14529 [gr-qc]} \BibitemShut
  {NoStop}%
\bibitem [{\citenamefont {Abbott}\ \emph
  {et~al.}(2021{\natexlab{d}})\citenamefont {Abbott} \emph
  {et~al.}}]{LIGOScientific:2021sio}%
  \BibitemOpen
  \bibfield  {author} {\bibinfo {author} {\bibfnamefont {R.}~\bibnamefont
  {Abbott}} \emph {et~al.} (\bibinfo {collaboration} {LIGO Scientific, VIRGO,
  KAGRA}),\ }\href@noop {} {\  (\bibinfo {year} {2021}{\natexlab{d}})},\
  \Eprint {http://arxiv.org/abs/2112.06861} {arXiv:2112.06861 [gr-qc]}
  \BibitemShut {NoStop}%
\bibitem [{\citenamefont {Barausse}\ \emph {et~al.}(2014)\citenamefont
  {Barausse}, \citenamefont {Cardoso},\ and\ \citenamefont
  {Pani}}]{Barausse:2014tra}%
  \BibitemOpen
  \bibfield  {author} {\bibinfo {author} {\bibfnamefont {E.}~\bibnamefont
  {Barausse}}, \bibinfo {author} {\bibfnamefont {V.}~\bibnamefont {Cardoso}}, \
  and\ \bibinfo {author} {\bibfnamefont {P.}~\bibnamefont {Pani}},\ }\href
  {\doibase 10.1103/PhysRevD.89.104059} {\bibfield  {journal} {\bibinfo
  {journal} {Phys. Rev. D}\ }\textbf {\bibinfo {volume} {89}},\ \bibinfo
  {pages} {104059} (\bibinfo {year} {2014})},\ \Eprint
  {http://arxiv.org/abs/1404.7149} {arXiv:1404.7149 [gr-qc]} \BibitemShut
  {NoStop}%
\bibitem [{\citenamefont {Cardoso}\ \emph
  {et~al.}(2022{\natexlab{a}})\citenamefont {Cardoso}, \citenamefont
  {Destounis}, \citenamefont {Duque}, \citenamefont {Macedo},\ and\
  \citenamefont {Maselli}}]{Cardoso:2021wlq}%
  \BibitemOpen
  \bibfield  {author} {\bibinfo {author} {\bibfnamefont {V.}~\bibnamefont
  {Cardoso}}, \bibinfo {author} {\bibfnamefont {K.}~\bibnamefont {Destounis}},
  \bibinfo {author} {\bibfnamefont {F.}~\bibnamefont {Duque}}, \bibinfo
  {author} {\bibfnamefont {R.~P.}\ \bibnamefont {Macedo}}, \ and\ \bibinfo
  {author} {\bibfnamefont {A.}~\bibnamefont {Maselli}},\ }\href {\doibase
  10.1103/PhysRevD.105.L061501} {\bibfield  {journal} {\bibinfo  {journal}
  {Phys. Rev. D}\ }\textbf {\bibinfo {volume} {105}},\ \bibinfo {pages}
  {L061501} (\bibinfo {year} {2022}{\natexlab{a}})},\ \Eprint
  {http://arxiv.org/abs/2109.00005} {arXiv:2109.00005 [gr-qc]} \BibitemShut
  {NoStop}%
\bibitem [{\citenamefont {Cardoso}\ \emph
  {et~al.}(2022{\natexlab{b}})\citenamefont {Cardoso}, \citenamefont
  {Destounis}, \citenamefont {Duque}, \citenamefont {Panosso~Macedo},\ and\
  \citenamefont {Maselli}}]{Cardoso:2022whc}%
  \BibitemOpen
  \bibfield  {author} {\bibinfo {author} {\bibfnamefont {V.}~\bibnamefont
  {Cardoso}}, \bibinfo {author} {\bibfnamefont {K.}~\bibnamefont {Destounis}},
  \bibinfo {author} {\bibfnamefont {F.}~\bibnamefont {Duque}}, \bibinfo
  {author} {\bibfnamefont {R.}~\bibnamefont {Panosso~Macedo}}, \ and\ \bibinfo
  {author} {\bibfnamefont {A.}~\bibnamefont {Maselli}},\ }\href {\doibase
  10.1103/PhysRevLett.129.241103} {\bibfield  {journal} {\bibinfo  {journal}
  {Phys. Rev. Lett.}\ }\textbf {\bibinfo {volume} {129}},\ \bibinfo {pages}
  {241103} (\bibinfo {year} {2022}{\natexlab{b}})},\ \Eprint
  {http://arxiv.org/abs/2210.01133} {arXiv:2210.01133 [gr-qc]} \BibitemShut
  {NoStop}%
\bibitem [{\citenamefont {Cole}\ \emph {et~al.}(2023)\citenamefont {Cole},
  \citenamefont {Bertone}, \citenamefont {Coogan}, \citenamefont {Gaggero},
  \citenamefont {Karydas}, \citenamefont {Kavanagh}, \citenamefont {Spieksma},\
  and\ \citenamefont {Tomaselli}}]{Cole:2022yzw}%
  \BibitemOpen
  \bibfield  {author} {\bibinfo {author} {\bibfnamefont {P.~S.}\ \bibnamefont
  {Cole}}, \bibinfo {author} {\bibfnamefont {G.}~\bibnamefont {Bertone}},
  \bibinfo {author} {\bibfnamefont {A.}~\bibnamefont {Coogan}}, \bibinfo
  {author} {\bibfnamefont {D.}~\bibnamefont {Gaggero}}, \bibinfo {author}
  {\bibfnamefont {T.}~\bibnamefont {Karydas}}, \bibinfo {author} {\bibfnamefont
  {B.~J.}\ \bibnamefont {Kavanagh}}, \bibinfo {author} {\bibfnamefont
  {T.~F.~M.}\ \bibnamefont {Spieksma}}, \ and\ \bibinfo {author} {\bibfnamefont
  {G.~M.}\ \bibnamefont {Tomaselli}},\ }\href {\doibase
  10.1038/s41550-023-01990-2} {\bibfield  {journal} {\bibinfo  {journal}
  {Nature Astron.}\ }\textbf {\bibinfo {volume} {7}},\ \bibinfo {pages} {943}
  (\bibinfo {year} {2023})},\ \Eprint {http://arxiv.org/abs/2211.01362}
  {arXiv:2211.01362 [gr-qc]} \BibitemShut {NoStop}%
\bibitem [{\citenamefont {Caneva~Santoro}\ \emph {et~al.}(2023)\citenamefont
  {Caneva~Santoro}, \citenamefont {Roy}, \citenamefont {Vicente}, \citenamefont
  {Haney}, \citenamefont {Piccinni}, \citenamefont {Del~Pozzo},\ and\
  \citenamefont {Martinez}}]{CanevaSantoro:2023aol}%
  \BibitemOpen
  \bibfield  {author} {\bibinfo {author} {\bibfnamefont {G.}~\bibnamefont
  {Caneva~Santoro}}, \bibinfo {author} {\bibfnamefont {S.}~\bibnamefont {Roy}},
  \bibinfo {author} {\bibfnamefont {R.}~\bibnamefont {Vicente}}, \bibinfo
  {author} {\bibfnamefont {M.}~\bibnamefont {Haney}}, \bibinfo {author}
  {\bibfnamefont {O.~J.}\ \bibnamefont {Piccinni}}, \bibinfo {author}
  {\bibfnamefont {W.}~\bibnamefont {Del~Pozzo}}, \ and\ \bibinfo {author}
  {\bibfnamefont {M.}~\bibnamefont {Martinez}},\ }\href@noop {} {\  (\bibinfo
  {year} {2023})},\ \Eprint {http://arxiv.org/abs/2309.05061} {arXiv:2309.05061
  [gr-qc]} \BibitemShut {NoStop}%
\bibitem [{\citenamefont {Bertone}\ \emph {et~al.}(2005)\citenamefont
  {Bertone}, \citenamefont {Hooper},\ and\ \citenamefont
  {Silk}}]{Bertone:2004pz}%
  \BibitemOpen
  \bibfield  {author} {\bibinfo {author} {\bibfnamefont {G.}~\bibnamefont
  {Bertone}}, \bibinfo {author} {\bibfnamefont {D.}~\bibnamefont {Hooper}}, \
  and\ \bibinfo {author} {\bibfnamefont {J.}~\bibnamefont {Silk}},\ }\href
  {\doibase 10.1016/j.physrep.2004.08.031} {\bibfield  {journal} {\bibinfo
  {journal} {Phys. Rept.}\ }\textbf {\bibinfo {volume} {405}},\ \bibinfo
  {pages} {279} (\bibinfo {year} {2005})},\ \Eprint
  {http://arxiv.org/abs/hep-ph/0404175} {arXiv:hep-ph/0404175} \BibitemShut
  {NoStop}%
\bibitem [{\citenamefont {Bertone}\ and\ \citenamefont
  {Tait}(2018)}]{Bertone:2018krk}%
  \BibitemOpen
  \bibfield  {author} {\bibinfo {author} {\bibfnamefont {G.}~\bibnamefont
  {Bertone}}\ and\ \bibinfo {author} {\bibfnamefont {T.}~\bibnamefont {Tait},
  \bibfnamefont {M.~P.}},\ }\href {\doibase 10.1038/s41586-018-0542-z}
  {\bibfield  {journal} {\bibinfo  {journal} {Nature}\ }\textbf {\bibinfo
  {volume} {562}},\ \bibinfo {pages} {51} (\bibinfo {year} {2018})},\ \Eprint
  {http://arxiv.org/abs/1810.01668} {arXiv:1810.01668 [astro-ph.CO]}
  \BibitemShut {NoStop}%
\bibitem [{\citenamefont {Bertone}\ \emph {et~al.}(2020)\citenamefont {Bertone}
  \emph {et~al.}}]{Bertone:2019irm}%
  \BibitemOpen
  \bibfield  {author} {\bibinfo {author} {\bibfnamefont {G.}~\bibnamefont
  {Bertone}} \emph {et~al.},\ }\href {\doibase
  10.21468/SciPostPhysCore.3.2.007} {\bibfield  {journal} {\bibinfo  {journal}
  {SciPost Phys. Core}\ }\textbf {\bibinfo {volume} {3}},\ \bibinfo {pages}
  {007} (\bibinfo {year} {2020})},\ \Eprint {http://arxiv.org/abs/1907.10610}
  {arXiv:1907.10610 [astro-ph.CO]} \BibitemShut {NoStop}%
\bibitem [{\citenamefont {Brito}\ \emph {et~al.}(2015)\citenamefont {Brito},
  \citenamefont {Cardoso},\ and\ \citenamefont {Pani}}]{Brito:2015oca}%
  \BibitemOpen
  \bibfield  {author} {\bibinfo {author} {\bibfnamefont {R.}~\bibnamefont
  {Brito}}, \bibinfo {author} {\bibfnamefont {V.}~\bibnamefont {Cardoso}}, \
  and\ \bibinfo {author} {\bibfnamefont {P.}~\bibnamefont {Pani}},\ }\href
  {\doibase 10.1007/978-3-319-19000-6} {\bibfield  {journal} {\bibinfo
  {journal} {Lect. Notes Phys.}\ }\textbf {\bibinfo {volume} {906}},\ \bibinfo
  {pages} {pp.1} (\bibinfo {year} {2015})},\ \Eprint
  {http://arxiv.org/abs/1501.06570} {arXiv:1501.06570 [gr-qc]} \BibitemShut
  {NoStop}%
%%CITATION = ARXIV:1501.06570;%%
\bibitem [{\citenamefont {Eardley}\ and\ \citenamefont
  {Press}(1975)}]{Eardley:1975kp}%
  \BibitemOpen
  \bibfield  {author} {\bibinfo {author} {\bibfnamefont {D.~M.}\ \bibnamefont
  {Eardley}}\ and\ \bibinfo {author} {\bibfnamefont {W.~H.}\ \bibnamefont
  {Press}},\ }\href {\doibase 10.1146/annurev.aa.13.090175.002121} {\bibfield
  {journal} {\bibinfo  {journal} {Ann. Rev. Astron. Astrophys.}\ }\textbf
  {\bibinfo {volume} {13}},\ \bibinfo {pages} {381} (\bibinfo {year}
  {1975})}\BibitemShut {NoStop}%
\bibitem [{\citenamefont {Gibbons}(1975)}]{Gibbons:1975kk}%
  \BibitemOpen
  \bibfield  {author} {\bibinfo {author} {\bibfnamefont {G.~W.}\ \bibnamefont
  {Gibbons}},\ }\href {\doibase 10.1007/BF01609829} {\bibfield  {journal}
  {\bibinfo  {journal} {Commun. Math. Phys.}\ }\textbf {\bibinfo {volume}
  {44}},\ \bibinfo {pages} {245} (\bibinfo {year} {1975})}\BibitemShut
  {NoStop}%
\bibitem [{\citenamefont {Cardoso}\ \emph
  {et~al.}(2016{\natexlab{a}})\citenamefont {Cardoso}, \citenamefont {Macedo},
  \citenamefont {Pani},\ and\ \citenamefont {Ferrari}}]{Cardoso:2016olt}%
  \BibitemOpen
  \bibfield  {author} {\bibinfo {author} {\bibfnamefont {V.}~\bibnamefont
  {Cardoso}}, \bibinfo {author} {\bibfnamefont {C.~F.~B.}\ \bibnamefont
  {Macedo}}, \bibinfo {author} {\bibfnamefont {P.}~\bibnamefont {Pani}}, \ and\
  \bibinfo {author} {\bibfnamefont {V.}~\bibnamefont {Ferrari}},\ }\href
  {\doibase 10.1088/1475-7516/2016/05/054} {\bibfield  {journal} {\bibinfo
  {journal} {JCAP}\ }\textbf {\bibinfo {volume} {05}},\ \bibinfo {pages} {054}
  (\bibinfo {year} {2016}{\natexlab{a}})},\ \bibinfo {note} {[Erratum: JCAP 04,
  E01 (2020)]},\ \Eprint {http://arxiv.org/abs/1604.07845} {arXiv:1604.07845
  [hep-ph]} \BibitemShut {NoStop}%
\bibitem [{\citenamefont {de~Freitas~Pacheco}\ \emph
  {et~al.}(2023)\citenamefont {de~Freitas~Pacheco}, \citenamefont {Kiritsis},
  \citenamefont {Lucca},\ and\ \citenamefont
  {Silk}}]{deFreitasPacheco:2023hpb}%
  \BibitemOpen
  \bibfield  {author} {\bibinfo {author} {\bibfnamefont {J.~A.}\ \bibnamefont
  {de~Freitas~Pacheco}}, \bibinfo {author} {\bibfnamefont {E.}~\bibnamefont
  {Kiritsis}}, \bibinfo {author} {\bibfnamefont {M.}~\bibnamefont {Lucca}}, \
  and\ \bibinfo {author} {\bibfnamefont {J.}~\bibnamefont {Silk}},\ }\href
  {\doibase 10.1103/PhysRevD.107.123525} {\bibfield  {journal} {\bibinfo
  {journal} {Phys. Rev. D}\ }\textbf {\bibinfo {volume} {107}},\ \bibinfo
  {pages} {123525} (\bibinfo {year} {2023})},\ \Eprint
  {http://arxiv.org/abs/2301.13215} {arXiv:2301.13215 [astro-ph.CO]}
  \BibitemShut {NoStop}%
\bibitem [{\citenamefont {Alonso-Monsalve}\ and\ \citenamefont
  {Kaiser}(2023{\natexlab{a}})}]{Alonso-Monsalve:2023brx}%
  \BibitemOpen
  \bibfield  {author} {\bibinfo {author} {\bibfnamefont {E.}~\bibnamefont
  {Alonso-Monsalve}}\ and\ \bibinfo {author} {\bibfnamefont {D.~I.}\
  \bibnamefont {Kaiser}},\ }\href@noop {} {\  (\bibinfo {year}
  {2023}{\natexlab{a}})},\ \Eprint {http://arxiv.org/abs/2310.16877}
  {arXiv:2310.16877 [hep-ph]} \BibitemShut {NoStop}%
\bibitem [{\citenamefont {Bai}\ and\ \citenamefont
  {Orlofsky}(2020)}]{Bai:2019zcd}%
  \BibitemOpen
  \bibfield  {author} {\bibinfo {author} {\bibfnamefont {Y.}~\bibnamefont
  {Bai}}\ and\ \bibinfo {author} {\bibfnamefont {N.}~\bibnamefont {Orlofsky}},\
  }\href {\doibase 10.1103/PhysRevD.101.055006} {\bibfield  {journal} {\bibinfo
   {journal} {Phys. Rev. D}\ }\textbf {\bibinfo {volume} {101}},\ \bibinfo
  {pages} {055006} (\bibinfo {year} {2020})},\ \Eprint
  {http://arxiv.org/abs/1906.04858} {arXiv:1906.04858 [hep-ph]} \BibitemShut
  {NoStop}%
\bibitem [{\citenamefont {Kritos}\ and\ \citenamefont
  {Silk}(2022)}]{Kritos:2021nsf}%
  \BibitemOpen
  \bibfield  {author} {\bibinfo {author} {\bibfnamefont {K.}~\bibnamefont
  {Kritos}}\ and\ \bibinfo {author} {\bibfnamefont {J.}~\bibnamefont {Silk}},\
  }\href {\doibase 10.1103/PhysRevD.105.063011} {\bibfield  {journal} {\bibinfo
   {journal} {Phys. Rev. D}\ }\textbf {\bibinfo {volume} {105}},\ \bibinfo
  {pages} {063011} (\bibinfo {year} {2022})},\ \Eprint
  {http://arxiv.org/abs/2109.09769} {arXiv:2109.09769 [gr-qc]} \BibitemShut
  {NoStop}%
\bibitem [{\citenamefont {De~Rujula}\ \emph {et~al.}(1990)\citenamefont
  {De~Rujula}, \citenamefont {Glashow},\ and\ \citenamefont
  {Sarid}}]{DeRujula:1989fe}%
  \BibitemOpen
  \bibfield  {author} {\bibinfo {author} {\bibfnamefont {A.}~\bibnamefont
  {De~Rujula}}, \bibinfo {author} {\bibfnamefont {S.~L.}\ \bibnamefont
  {Glashow}}, \ and\ \bibinfo {author} {\bibfnamefont {U.}~\bibnamefont
  {Sarid}},\ }\href {\doibase 10.1016/0550-3213(90)90227-5} {\bibfield
  {journal} {\bibinfo  {journal} {Nucl. Phys. B}\ }\textbf {\bibinfo {volume}
  {333}},\ \bibinfo {pages} {173} (\bibinfo {year} {1990})}\BibitemShut
  {NoStop}%
\bibitem [{\citenamefont {Holdom}(1986)}]{Holdom:1985ag}%
  \BibitemOpen
  \bibfield  {author} {\bibinfo {author} {\bibfnamefont {B.}~\bibnamefont
  {Holdom}},\ }\href {\doibase 10.1016/0370-2693(86)91377-8} {\bibfield
  {journal} {\bibinfo  {journal} {Phys. Lett. B}\ }\textbf {\bibinfo {volume}
  {166}},\ \bibinfo {pages} {196} (\bibinfo {year} {1986})}\BibitemShut
  {NoStop}%
\bibitem [{\citenamefont {Davidson}\ \emph {et~al.}(2000)\citenamefont
  {Davidson}, \citenamefont {Hannestad},\ and\ \citenamefont
  {Raffelt}}]{Davidson:2000hf}%
  \BibitemOpen
  \bibfield  {author} {\bibinfo {author} {\bibfnamefont {S.}~\bibnamefont
  {Davidson}}, \bibinfo {author} {\bibfnamefont {S.}~\bibnamefont {Hannestad}},
  \ and\ \bibinfo {author} {\bibfnamefont {G.}~\bibnamefont {Raffelt}},\ }\href
  {\doibase 10.1088/1126-6708/2000/05/003} {\bibfield  {journal} {\bibinfo
  {journal} {JHEP}\ }\textbf {\bibinfo {volume} {05}},\ \bibinfo {pages} {003}
  (\bibinfo {year} {2000})},\ \Eprint {http://arxiv.org/abs/hep-ph/0001179}
  {arXiv:hep-ph/0001179} \BibitemShut {NoStop}%
\bibitem [{\citenamefont {Dubovsky}\ \emph {et~al.}(2004)\citenamefont
  {Dubovsky}, \citenamefont {Gorbunov},\ and\ \citenamefont
  {Rubtsov}}]{Dubovsky:2003yn}%
  \BibitemOpen
  \bibfield  {author} {\bibinfo {author} {\bibfnamefont {S.~L.}\ \bibnamefont
  {Dubovsky}}, \bibinfo {author} {\bibfnamefont {D.~S.}\ \bibnamefont
  {Gorbunov}}, \ and\ \bibinfo {author} {\bibfnamefont {G.~I.}\ \bibnamefont
  {Rubtsov}},\ }\href {\doibase 10.1134/1.1675909} {\bibfield  {journal}
  {\bibinfo  {journal} {JETP Lett.}\ }\textbf {\bibinfo {volume} {79}},\
  \bibinfo {pages} {1} (\bibinfo {year} {2004})},\ \Eprint
  {http://arxiv.org/abs/hep-ph/0311189} {arXiv:hep-ph/0311189} \BibitemShut
  {NoStop}%
\bibitem [{\citenamefont {Sigurdson}\ \emph {et~al.}(2004)\citenamefont
  {Sigurdson}, \citenamefont {Doran}, \citenamefont {Kurylov}, \citenamefont
  {Caldwell},\ and\ \citenamefont {Kamionkowski}}]{Sigurdson:2004zp}%
  \BibitemOpen
  \bibfield  {author} {\bibinfo {author} {\bibfnamefont {K.}~\bibnamefont
  {Sigurdson}}, \bibinfo {author} {\bibfnamefont {M.}~\bibnamefont {Doran}},
  \bibinfo {author} {\bibfnamefont {A.}~\bibnamefont {Kurylov}}, \bibinfo
  {author} {\bibfnamefont {R.~R.}\ \bibnamefont {Caldwell}}, \ and\ \bibinfo
  {author} {\bibfnamefont {M.}~\bibnamefont {Kamionkowski}},\ }\href {\doibase
  10.1103/PhysRevD.70.083501} {\bibfield  {journal} {\bibinfo  {journal} {Phys.
  Rev. D}\ }\textbf {\bibinfo {volume} {70}},\ \bibinfo {pages} {083501}
  (\bibinfo {year} {2004})},\ \bibinfo {note} {[Erratum: Phys.Rev.D 73, 089903
  (2006)]},\ \Eprint {http://arxiv.org/abs/astro-ph/0406355}
  {arXiv:astro-ph/0406355} \BibitemShut {NoStop}%
\bibitem [{\citenamefont {Gies}\ \emph
  {et~al.}(2006{\natexlab{a}})\citenamefont {Gies}, \citenamefont {Jaeckel},\
  and\ \citenamefont {Ringwald}}]{Gies:2006ca}%
  \BibitemOpen
  \bibfield  {author} {\bibinfo {author} {\bibfnamefont {H.}~\bibnamefont
  {Gies}}, \bibinfo {author} {\bibfnamefont {J.}~\bibnamefont {Jaeckel}}, \
  and\ \bibinfo {author} {\bibfnamefont {A.}~\bibnamefont {Ringwald}},\ }\href
  {\doibase 10.1103/PhysRevLett.97.140402} {\bibfield  {journal} {\bibinfo
  {journal} {Phys. Rev. Lett.}\ }\textbf {\bibinfo {volume} {97}},\ \bibinfo
  {pages} {140402} (\bibinfo {year} {2006}{\natexlab{a}})},\ \Eprint
  {http://arxiv.org/abs/hep-ph/0607118} {arXiv:hep-ph/0607118} \BibitemShut
  {NoStop}%
\bibitem [{\citenamefont {Gies}\ \emph
  {et~al.}(2006{\natexlab{b}})\citenamefont {Gies}, \citenamefont {Jaeckel},\
  and\ \citenamefont {Ringwald}}]{Gies:2006hv}%
  \BibitemOpen
  \bibfield  {author} {\bibinfo {author} {\bibfnamefont {H.}~\bibnamefont
  {Gies}}, \bibinfo {author} {\bibfnamefont {J.}~\bibnamefont {Jaeckel}}, \
  and\ \bibinfo {author} {\bibfnamefont {A.}~\bibnamefont {Ringwald}},\ }\href
  {\doibase 10.1209/epl/i2006-10356-5} {\bibfield  {journal} {\bibinfo
  {journal} {EPL}\ }\textbf {\bibinfo {volume} {76}},\ \bibinfo {pages} {794}
  (\bibinfo {year} {2006}{\natexlab{b}})},\ \Eprint
  {http://arxiv.org/abs/hep-ph/0608238} {arXiv:hep-ph/0608238} \BibitemShut
  {NoStop}%
\bibitem [{\citenamefont {Burrage}\ \emph {et~al.}(2009)\citenamefont
  {Burrage}, \citenamefont {Jaeckel}, \citenamefont {Redondo},\ and\
  \citenamefont {Ringwald}}]{Burrage:2009yz}%
  \BibitemOpen
  \bibfield  {author} {\bibinfo {author} {\bibfnamefont {C.}~\bibnamefont
  {Burrage}}, \bibinfo {author} {\bibfnamefont {J.}~\bibnamefont {Jaeckel}},
  \bibinfo {author} {\bibfnamefont {J.}~\bibnamefont {Redondo}}, \ and\
  \bibinfo {author} {\bibfnamefont {A.}~\bibnamefont {Ringwald}},\ }\href
  {\doibase 10.1088/1475-7516/2009/11/002} {\bibfield  {journal} {\bibinfo
  {journal} {JCAP}\ }\textbf {\bibinfo {volume} {11}},\ \bibinfo {pages} {002}
  (\bibinfo {year} {2009})},\ \Eprint {http://arxiv.org/abs/0909.0649}
  {arXiv:0909.0649 [astro-ph.CO]} \BibitemShut {NoStop}%
\bibitem [{\citenamefont {Ahlers}(2009)}]{Ahlers:2009kh}%
  \BibitemOpen
  \bibfield  {author} {\bibinfo {author} {\bibfnamefont {M.}~\bibnamefont
  {Ahlers}},\ }\href {\doibase 10.1103/PhysRevD.80.023513} {\bibfield
  {journal} {\bibinfo  {journal} {Phys. Rev. D}\ }\textbf {\bibinfo {volume}
  {80}},\ \bibinfo {pages} {023513} (\bibinfo {year} {2009})},\ \Eprint
  {http://arxiv.org/abs/0904.0998} {arXiv:0904.0998 [hep-ph]} \BibitemShut
  {NoStop}%
\bibitem [{\citenamefont {McDermott}\ \emph {et~al.}(2011)\citenamefont
  {McDermott}, \citenamefont {Yu},\ and\ \citenamefont
  {Zurek}}]{McDermott:2010pa}%
  \BibitemOpen
  \bibfield  {author} {\bibinfo {author} {\bibfnamefont {S.~D.}\ \bibnamefont
  {McDermott}}, \bibinfo {author} {\bibfnamefont {H.-B.}\ \bibnamefont {Yu}}, \
  and\ \bibinfo {author} {\bibfnamefont {K.~M.}\ \bibnamefont {Zurek}},\ }\href
  {\doibase 10.1103/PhysRevD.83.063509} {\bibfield  {journal} {\bibinfo
  {journal} {Phys. Rev. D}\ }\textbf {\bibinfo {volume} {83}},\ \bibinfo
  {pages} {063509} (\bibinfo {year} {2011})},\ \Eprint
  {http://arxiv.org/abs/1011.2907} {arXiv:1011.2907 [hep-ph]} \BibitemShut
  {NoStop}%
\bibitem [{\citenamefont {Dolgov}\ \emph {et~al.}(2013)\citenamefont {Dolgov},
  \citenamefont {Dubovsky}, \citenamefont {Rubtsov},\ and\ \citenamefont
  {Tkachev}}]{Dolgov:2013una}%
  \BibitemOpen
  \bibfield  {author} {\bibinfo {author} {\bibfnamefont {A.~D.}\ \bibnamefont
  {Dolgov}}, \bibinfo {author} {\bibfnamefont {S.~L.}\ \bibnamefont
  {Dubovsky}}, \bibinfo {author} {\bibfnamefont {G.~I.}\ \bibnamefont
  {Rubtsov}}, \ and\ \bibinfo {author} {\bibfnamefont {I.~I.}\ \bibnamefont
  {Tkachev}},\ }\href {\doibase 10.1103/PhysRevD.88.117701} {\bibfield
  {journal} {\bibinfo  {journal} {Phys. Rev. D}\ }\textbf {\bibinfo {volume}
  {88}},\ \bibinfo {pages} {117701} (\bibinfo {year} {2013})},\ \Eprint
  {http://arxiv.org/abs/1310.2376} {arXiv:1310.2376 [hep-ph]} \BibitemShut
  {NoStop}%
\bibitem [{\citenamefont {Haas}\ \emph {et~al.}(2015)\citenamefont {Haas},
  \citenamefont {Hill}, \citenamefont {Izaguirre},\ and\ \citenamefont
  {Yavin}}]{Haas:2014dda}%
  \BibitemOpen
  \bibfield  {author} {\bibinfo {author} {\bibfnamefont {A.}~\bibnamefont
  {Haas}}, \bibinfo {author} {\bibfnamefont {C.~S.}\ \bibnamefont {Hill}},
  \bibinfo {author} {\bibfnamefont {E.}~\bibnamefont {Izaguirre}}, \ and\
  \bibinfo {author} {\bibfnamefont {I.}~\bibnamefont {Yavin}},\ }\href
  {\doibase 10.1016/j.physletb.2015.04.062} {\bibfield  {journal} {\bibinfo
  {journal} {Phys. Lett. B}\ }\textbf {\bibinfo {volume} {746}},\ \bibinfo
  {pages} {117} (\bibinfo {year} {2015})},\ \Eprint
  {http://arxiv.org/abs/1410.6816} {arXiv:1410.6816 [hep-ph]} \BibitemShut
  {NoStop}%
\bibitem [{\citenamefont {Khalil}\ \emph {et~al.}(2018)\citenamefont {Khalil},
  \citenamefont {Sennett}, \citenamefont {Steinhoff}, \citenamefont {Vines},\
  and\ \citenamefont {Buonanno}}]{Khalil:2018aaj}%
  \BibitemOpen
  \bibfield  {author} {\bibinfo {author} {\bibfnamefont {M.}~\bibnamefont
  {Khalil}}, \bibinfo {author} {\bibfnamefont {N.}~\bibnamefont {Sennett}},
  \bibinfo {author} {\bibfnamefont {J.}~\bibnamefont {Steinhoff}}, \bibinfo
  {author} {\bibfnamefont {J.}~\bibnamefont {Vines}}, \ and\ \bibinfo {author}
  {\bibfnamefont {A.}~\bibnamefont {Buonanno}},\ }\href {\doibase
  10.1103/PhysRevD.98.104010} {\bibfield  {journal} {\bibinfo  {journal} {Phys.
  Rev. D}\ }\textbf {\bibinfo {volume} {98}},\ \bibinfo {pages} {104010}
  (\bibinfo {year} {2018})},\ \Eprint {http://arxiv.org/abs/1809.03109}
  {arXiv:1809.03109 [gr-qc]} \BibitemShut {NoStop}%
\bibitem [{\citenamefont {Caputo}\ \emph {et~al.}(2019)\citenamefont {Caputo},
  \citenamefont {Sberna}, \citenamefont {Frias}, \citenamefont {Blas},
  \citenamefont {Pani}, \citenamefont {Shao},\ and\ \citenamefont
  {Yan}}]{Caputo:2019tms}%
  \BibitemOpen
  \bibfield  {author} {\bibinfo {author} {\bibfnamefont {A.}~\bibnamefont
  {Caputo}}, \bibinfo {author} {\bibfnamefont {L.}~\bibnamefont {Sberna}},
  \bibinfo {author} {\bibfnamefont {M.}~\bibnamefont {Frias}}, \bibinfo
  {author} {\bibfnamefont {D.}~\bibnamefont {Blas}}, \bibinfo {author}
  {\bibfnamefont {P.}~\bibnamefont {Pani}}, \bibinfo {author} {\bibfnamefont
  {L.}~\bibnamefont {Shao}}, \ and\ \bibinfo {author} {\bibfnamefont
  {W.}~\bibnamefont {Yan}},\ }\href {\doibase 10.1103/PhysRevD.100.063515}
  {\bibfield  {journal} {\bibinfo  {journal} {Phys. Rev. D}\ }\textbf {\bibinfo
  {volume} {100}},\ \bibinfo {pages} {063515} (\bibinfo {year} {2019})},\
  \Eprint {http://arxiv.org/abs/1902.02695} {arXiv:1902.02695 [astro-ph.CO]}
  \BibitemShut {NoStop}%
\bibitem [{\citenamefont {Gupta}\ \emph {et~al.}(2021)\citenamefont {Gupta},
  \citenamefont {Spieksma}, \citenamefont {Pang}, \citenamefont {Koekoek},\
  and\ \citenamefont {Broeck}}]{Gupta:2021rod}%
  \BibitemOpen
  \bibfield  {author} {\bibinfo {author} {\bibfnamefont {P.~K.}\ \bibnamefont
  {Gupta}}, \bibinfo {author} {\bibfnamefont {T.~F.~M.}\ \bibnamefont
  {Spieksma}}, \bibinfo {author} {\bibfnamefont {P.~T.~H.}\ \bibnamefont
  {Pang}}, \bibinfo {author} {\bibfnamefont {G.}~\bibnamefont {Koekoek}}, \
  and\ \bibinfo {author} {\bibfnamefont {C.~V.~D.}\ \bibnamefont {Broeck}},\
  }\href {\doibase 10.1103/PhysRevD.104.063041} {\bibfield  {journal} {\bibinfo
   {journal} {Phys. Rev. D}\ }\textbf {\bibinfo {volume} {104}},\ \bibinfo
  {pages} {063041} (\bibinfo {year} {2021})},\ \Eprint
  {http://arxiv.org/abs/2107.12111} {arXiv:2107.12111 [gr-qc]} \BibitemShut
  {NoStop}%
\bibitem [{\citenamefont {Carullo}\ \emph {et~al.}(2022)\citenamefont
  {Carullo}, \citenamefont {Laghi}, \citenamefont {Johnson-McDaniel},
  \citenamefont {Del~Pozzo}, \citenamefont {Dias}, \citenamefont {Godazgar},\
  and\ \citenamefont {Santos}}]{Carullo:2021oxn}%
  \BibitemOpen
  \bibfield  {author} {\bibinfo {author} {\bibfnamefont {G.}~\bibnamefont
  {Carullo}}, \bibinfo {author} {\bibfnamefont {D.}~\bibnamefont {Laghi}},
  \bibinfo {author} {\bibfnamefont {N.~K.}\ \bibnamefont {Johnson-McDaniel}},
  \bibinfo {author} {\bibfnamefont {W.}~\bibnamefont {Del~Pozzo}}, \bibinfo
  {author} {\bibfnamefont {O.~J.~C.}\ \bibnamefont {Dias}}, \bibinfo {author}
  {\bibfnamefont {M.}~\bibnamefont {Godazgar}}, \ and\ \bibinfo {author}
  {\bibfnamefont {J.~E.}\ \bibnamefont {Santos}},\ }\href {\doibase
  10.1103/PhysRevD.105.062009} {\bibfield  {journal} {\bibinfo  {journal}
  {Phys. Rev. D}\ }\textbf {\bibinfo {volume} {105}},\ \bibinfo {pages}
  {062009} (\bibinfo {year} {2022})},\ \Eprint
  {http://arxiv.org/abs/2109.13961} {arXiv:2109.13961 [gr-qc]} \BibitemShut
  {NoStop}%
\bibitem [{\citenamefont {Fiorillo}\ and\ \citenamefont
  {Vitagliano}(2024)}]{Fiorillo:2024upk}%
  \BibitemOpen
  \bibfield  {author} {\bibinfo {author} {\bibfnamefont {D.~F.~G.}\
  \bibnamefont {Fiorillo}}\ and\ \bibinfo {author} {\bibfnamefont
  {E.}~\bibnamefont {Vitagliano}},\ }\href@noop {} {\  (\bibinfo {year}
  {2024})},\ \Eprint {http://arxiv.org/abs/2404.07714} {arXiv:2404.07714
  [hep-ph]} \BibitemShut {NoStop}%
\bibitem [{\citenamefont {Bah}\ and\ \citenamefont
  {Heidmann}(2021{\natexlab{a}})}]{Bah:2020ogh}%
  \BibitemOpen
  \bibfield  {author} {\bibinfo {author} {\bibfnamefont {I.}~\bibnamefont
  {Bah}}\ and\ \bibinfo {author} {\bibfnamefont {P.}~\bibnamefont {Heidmann}},\
  }\href {\doibase 10.1103/PhysRevLett.126.151101} {\bibfield  {journal}
  {\bibinfo  {journal} {Phys. Rev. Lett.}\ }\textbf {\bibinfo {volume} {126}},\
  \bibinfo {pages} {151101} (\bibinfo {year} {2021}{\natexlab{a}})},\ \Eprint
  {http://arxiv.org/abs/2011.08851} {arXiv:2011.08851 [hep-th]} \BibitemShut
  {NoStop}%
\bibitem [{\citenamefont {Bah}\ and\ \citenamefont
  {Heidmann}(2021{\natexlab{b}})}]{Bah:2021owp}%
  \BibitemOpen
  \bibfield  {author} {\bibinfo {author} {\bibfnamefont {I.}~\bibnamefont
  {Bah}}\ and\ \bibinfo {author} {\bibfnamefont {P.}~\bibnamefont {Heidmann}},\
  }\href {\doibase 10.1007/JHEP09(2021)128} {\bibfield  {journal} {\bibinfo
  {journal} {JHEP}\ }\textbf {\bibinfo {volume} {09}},\ \bibinfo {pages} {128}
  (\bibinfo {year} {2021}{\natexlab{b}})},\ \Eprint
  {http://arxiv.org/abs/2106.05118} {arXiv:2106.05118 [hep-th]} \BibitemShut
  {NoStop}%
\bibitem [{\citenamefont {Cardoso}\ \emph
  {et~al.}(2021{\natexlab{a}})\citenamefont {Cardoso}, \citenamefont {Guo},
  \citenamefont {Macedo},\ and\ \citenamefont {Pani}}]{Cardoso:2020nst}%
  \BibitemOpen
  \bibfield  {author} {\bibinfo {author} {\bibfnamefont {V.}~\bibnamefont
  {Cardoso}}, \bibinfo {author} {\bibfnamefont {W.-D.}\ \bibnamefont {Guo}},
  \bibinfo {author} {\bibfnamefont {C.~F.~B.}\ \bibnamefont {Macedo}}, \ and\
  \bibinfo {author} {\bibfnamefont {P.}~\bibnamefont {Pani}},\ }\href {\doibase
  10.1093/mnras/stab404} {\bibfield  {journal} {\bibinfo  {journal} {Mon. Not.
  Roy. Astron. Soc.}\ }\textbf {\bibinfo {volume} {503}},\ \bibinfo {pages}
  {563} (\bibinfo {year} {2021}{\natexlab{a}})},\ \Eprint
  {http://arxiv.org/abs/2009.07287} {arXiv:2009.07287 [gr-qc]} \BibitemShut
  {NoStop}%
\bibitem [{\citenamefont {Leaver}(1990)}]{Leaver:1990zz}%
  \BibitemOpen
  \bibfield  {author} {\bibinfo {author} {\bibfnamefont {E.~W.}\ \bibnamefont
  {Leaver}},\ }\href {\doibase 10.1103/PhysRevD.41.2986} {\bibfield  {journal}
  {\bibinfo  {journal} {Phys. Rev. D}\ }\textbf {\bibinfo {volume} {41}},\
  \bibinfo {pages} {2986} (\bibinfo {year} {1990})}\BibitemShut {NoStop}%
\bibitem [{\citenamefont {Berti}\ \emph {et~al.}(2009)\citenamefont {Berti},
  \citenamefont {Cardoso},\ and\ \citenamefont {Starinets}}]{Berti:2009kk}%
  \BibitemOpen
  \bibfield  {author} {\bibinfo {author} {\bibfnamefont {E.}~\bibnamefont
  {Berti}}, \bibinfo {author} {\bibfnamefont {V.}~\bibnamefont {Cardoso}}, \
  and\ \bibinfo {author} {\bibfnamefont {A.~O.}\ \bibnamefont {Starinets}},\
  }\href {\doibase 10.1088/0264-9381/26/16/163001} {\bibfield  {journal}
  {\bibinfo  {journal} {Class. Quant. Grav.}\ }\textbf {\bibinfo {volume}
  {26}},\ \bibinfo {pages} {163001} (\bibinfo {year} {2009})},\ \Eprint
  {http://arxiv.org/abs/0905.2975} {arXiv:0905.2975 [gr-qc]} \BibitemShut
  {NoStop}%
%%CITATION = ARXIV:0905.2975;%%
\bibitem [{\citenamefont {Cannizzaro}\ \emph
  {et~al.}(2021{\natexlab{a}})\citenamefont {Cannizzaro}, \citenamefont
  {Caputo}, \citenamefont {Sberna},\ and\ \citenamefont
  {Pani}}]{Cannizzaro:2020uap}%
  \BibitemOpen
  \bibfield  {author} {\bibinfo {author} {\bibfnamefont {E.}~\bibnamefont
  {Cannizzaro}}, \bibinfo {author} {\bibfnamefont {A.}~\bibnamefont {Caputo}},
  \bibinfo {author} {\bibfnamefont {L.}~\bibnamefont {Sberna}}, \ and\ \bibinfo
  {author} {\bibfnamefont {P.}~\bibnamefont {Pani}},\ }\href {\doibase
  10.1103/PhysRevD.103.124018} {\bibfield  {journal} {\bibinfo  {journal}
  {Phys. Rev. D}\ }\textbf {\bibinfo {volume} {103}},\ \bibinfo {pages}
  {124018} (\bibinfo {year} {2021}{\natexlab{a}})},\ \Eprint
  {http://arxiv.org/abs/2012.05114} {arXiv:2012.05114 [gr-qc]} \BibitemShut
  {NoStop}%
\bibitem [{\citenamefont {Cannizzaro}\ \emph
  {et~al.}(2021{\natexlab{b}})\citenamefont {Cannizzaro}, \citenamefont
  {Caputo}, \citenamefont {Sberna},\ and\ \citenamefont
  {Pani}}]{Cannizzaro:2021zbp}%
  \BibitemOpen
  \bibfield  {author} {\bibinfo {author} {\bibfnamefont {E.}~\bibnamefont
  {Cannizzaro}}, \bibinfo {author} {\bibfnamefont {A.}~\bibnamefont {Caputo}},
  \bibinfo {author} {\bibfnamefont {L.}~\bibnamefont {Sberna}}, \ and\ \bibinfo
  {author} {\bibfnamefont {P.}~\bibnamefont {Pani}},\ }\href {\doibase
  10.1103/PhysRevD.104.104048} {\bibfield  {journal} {\bibinfo  {journal}
  {Phys. Rev. D}\ }\textbf {\bibinfo {volume} {104}},\ \bibinfo {pages}
  {104048} (\bibinfo {year} {2021}{\natexlab{b}})},\ \Eprint
  {http://arxiv.org/abs/2107.01174} {arXiv:2107.01174 [gr-qc]} \BibitemShut
  {NoStop}%
\bibitem [{\citenamefont {Einstein}(1939)}]{Einstein:1939ms}%
  \BibitemOpen
  \bibfield  {author} {\bibinfo {author} {\bibfnamefont {A.}~\bibnamefont
  {Einstein}},\ }\href {\doibase 10.2307/1968902} {\bibfield  {journal}
  {\bibinfo  {journal} {Annals Math.}\ }\textbf {\bibinfo {volume} {40}},\
  \bibinfo {pages} {922} (\bibinfo {year} {1939})}\BibitemShut {NoStop}%
\bibitem [{\citenamefont {{Geralico}}\ \emph {et~al.}(2012)\citenamefont
  {{Geralico}}, \citenamefont {{Pompi}},\ and\ \citenamefont
  {{Ruffini}}}]{2012IJMPS..12..146G}%
  \BibitemOpen
  \bibfield  {author} {\bibinfo {author} {\bibfnamefont {A.}~\bibnamefont
  {{Geralico}}}, \bibinfo {author} {\bibfnamefont {F.}~\bibnamefont {{Pompi}}},
  \ and\ \bibinfo {author} {\bibfnamefont {R.}~\bibnamefont {{Ruffini}}},\ }in\
  \href {\doibase 10.1142/S2010194512006356} {\emph {\bibinfo {booktitle}
  {International Journal of Modern Physics Conference Series}}},\ \bibinfo
  {series} {International Journal of Modern Physics Conference Series},
  Vol.~\bibinfo {volume} {12}\ (\bibinfo {year} {2012})\ pp.\ \bibinfo {pages}
  {146--173}\BibitemShut {NoStop}%
\bibitem [{\citenamefont {Feng}\ \emph {et~al.}(2023)\citenamefont {Feng},
  \citenamefont {Chakraborty},\ and\ \citenamefont {Cardoso}}]{Feng:2022evy}%
  \BibitemOpen
  \bibfield  {author} {\bibinfo {author} {\bibfnamefont {J.~C.}\ \bibnamefont
  {Feng}}, \bibinfo {author} {\bibfnamefont {S.}~\bibnamefont {Chakraborty}}, \
  and\ \bibinfo {author} {\bibfnamefont {V.}~\bibnamefont {Cardoso}},\ }\href
  {\doibase 10.1103/PhysRevD.107.044050} {\bibfield  {journal} {\bibinfo
  {journal} {Phys. Rev. D}\ }\textbf {\bibinfo {volume} {107}},\ \bibinfo
  {pages} {044050} (\bibinfo {year} {2023})},\ \Eprint
  {http://arxiv.org/abs/2211.05261} {arXiv:2211.05261 [gr-qc]} \BibitemShut
  {NoStop}%
\bibitem [{\citenamefont {Cannizzaro}\ \emph {et~al.}(2023)\citenamefont
  {Cannizzaro}, \citenamefont {Corelli},\ and\ \citenamefont
  {Pani}}]{Cannizzaro:2023ltu}%
  \BibitemOpen
  \bibfield  {author} {\bibinfo {author} {\bibfnamefont {E.}~\bibnamefont
  {Cannizzaro}}, \bibinfo {author} {\bibfnamefont {F.}~\bibnamefont {Corelli}},
  \ and\ \bibinfo {author} {\bibfnamefont {P.}~\bibnamefont {Pani}},\
  }\href@noop {} {\  (\bibinfo {year} {2023})},\ \Eprint
  {http://arxiv.org/abs/2306.12490} {arXiv:2306.12490 [gr-qc]} \BibitemShut
  {NoStop}%
\bibitem [{\citenamefont {Spieksma}\ \emph {et~al.}(2023)\citenamefont
  {Spieksma}, \citenamefont {Cannizzaro}, \citenamefont {Ikeda}, \citenamefont
  {Cardoso},\ and\ \citenamefont {Chen}}]{Spieksma:2023vwl}%
  \BibitemOpen
  \bibfield  {author} {\bibinfo {author} {\bibfnamefont {T.~F.~M.}\
  \bibnamefont {Spieksma}}, \bibinfo {author} {\bibfnamefont {E.}~\bibnamefont
  {Cannizzaro}}, \bibinfo {author} {\bibfnamefont {T.}~\bibnamefont {Ikeda}},
  \bibinfo {author} {\bibfnamefont {V.}~\bibnamefont {Cardoso}}, \ and\
  \bibinfo {author} {\bibfnamefont {Y.}~\bibnamefont {Chen}},\ }\href {\doibase
  10.1103/PhysRevD.108.063013} {\bibfield  {journal} {\bibinfo  {journal}
  {Phys. Rev. D}\ }\textbf {\bibinfo {volume} {108}},\ \bibinfo {pages}
  {063013} (\bibinfo {year} {2023})},\ \Eprint
  {http://arxiv.org/abs/2306.16447} {arXiv:2306.16447 [gr-qc]} \BibitemShut
  {NoStop}%
\bibitem [{\citenamefont {Armitage}\ and\ \citenamefont
  {Natarajan}(2005)}]{Armitage:2005xq}%
  \BibitemOpen
  \bibfield  {author} {\bibinfo {author} {\bibfnamefont {P.~J.}\ \bibnamefont
  {Armitage}}\ and\ \bibinfo {author} {\bibfnamefont {P.}~\bibnamefont
  {Natarajan}},\ }\href {\doibase 10.1086/497108} {\bibfield  {journal}
  {\bibinfo  {journal} {Astrophys. J.}\ }\textbf {\bibinfo {volume} {634}},\
  \bibinfo {pages} {921} (\bibinfo {year} {2005})},\ \Eprint
  {http://arxiv.org/abs/astro-ph/0508493} {arXiv:astro-ph/0508493} \BibitemShut
  {NoStop}%
\bibitem [{\citenamefont {Artymowicz}\ and\ \citenamefont
  {Lubow}(1994)}]{Artymowicz:1994bw}%
  \BibitemOpen
  \bibfield  {author} {\bibinfo {author} {\bibfnamefont {P.}~\bibnamefont
  {Artymowicz}}\ and\ \bibinfo {author} {\bibfnamefont {S.~H.}\ \bibnamefont
  {Lubow}},\ }\href {\doibase 10.1086/173679} {\bibfield  {journal} {\bibinfo
  {journal} {Astrophys. J.}\ }\textbf {\bibinfo {volume} {421}},\ \bibinfo
  {pages} {651} (\bibinfo {year} {1994})}\BibitemShut {NoStop}%
\bibitem [{\citenamefont {Gr\"obner}\ \emph {et~al.}(2020)\citenamefont
  {Gr\"obner}, \citenamefont {Ishibashi}, \citenamefont {Tiwari}, \citenamefont
  {Haney},\ and\ \citenamefont {Jetzer}}]{Grobner:2020drr}%
  \BibitemOpen
  \bibfield  {author} {\bibinfo {author} {\bibfnamefont {M.}~\bibnamefont
  {Gr\"obner}}, \bibinfo {author} {\bibfnamefont {W.}~\bibnamefont
  {Ishibashi}}, \bibinfo {author} {\bibfnamefont {S.}~\bibnamefont {Tiwari}},
  \bibinfo {author} {\bibfnamefont {M.}~\bibnamefont {Haney}}, \ and\ \bibinfo
  {author} {\bibfnamefont {P.}~\bibnamefont {Jetzer}},\ }\href {\doibase
  10.1051/0004-6361/202037681} {\bibfield  {journal} {\bibinfo  {journal}
  {Astron. Astrophys.}\ }\textbf {\bibinfo {volume} {638}},\ \bibinfo {pages}
  {A119} (\bibinfo {year} {2020})},\ \Eprint {http://arxiv.org/abs/2005.03571}
  {arXiv:2005.03571 [astro-ph.GA]} \BibitemShut {NoStop}%
\bibitem [{\citenamefont {Farris}\ \emph {et~al.}(2015)\citenamefont {Farris},
  \citenamefont {Duffell}, \citenamefont {MacFadyen},\ and\ \citenamefont
  {Haiman}}]{Farris:2014zjo}%
  \BibitemOpen
  \bibfield  {author} {\bibinfo {author} {\bibfnamefont {B.~D.}\ \bibnamefont
  {Farris}}, \bibinfo {author} {\bibfnamefont {P.}~\bibnamefont {Duffell}},
  \bibinfo {author} {\bibfnamefont {A.~I.}\ \bibnamefont {MacFadyen}}, \ and\
  \bibinfo {author} {\bibfnamefont {Z.}~\bibnamefont {Haiman}},\ }\href
  {\doibase 10.1093/mnrasl/slu184} {\bibfield  {journal} {\bibinfo  {journal}
  {Mon. Not. Roy. Astron. Soc.}\ }\textbf {\bibinfo {volume} {447}},\ \bibinfo
  {pages} {L80} (\bibinfo {year} {2015})},\ \Eprint
  {http://arxiv.org/abs/1409.5124} {arXiv:1409.5124 [astro-ph.HE]} \BibitemShut
  {NoStop}%
\bibitem [{\citenamefont {Ishibashi}\ and\ \citenamefont
  {Gr\"obner}(2020)}]{Ishibashi:2020zzy}%
  \BibitemOpen
  \bibfield  {author} {\bibinfo {author} {\bibfnamefont {W.}~\bibnamefont
  {Ishibashi}}\ and\ \bibinfo {author} {\bibfnamefont {M.}~\bibnamefont
  {Gr\"obner}},\ }\href {\doibase 10.1051/0004-6361/202037799} {\bibfield
  {journal} {\bibinfo  {journal} {Astron. Astrophys.}\ }\textbf {\bibinfo
  {volume} {639}},\ \bibinfo {pages} {A108} (\bibinfo {year} {2020})},\ \Eprint
  {http://arxiv.org/abs/2006.07407} {arXiv:2006.07407 [astro-ph.GA]}
  \BibitemShut {NoStop}%
\bibitem [{\citenamefont {{D'Orazio}}\ \emph {et~al.}(2013)\citenamefont
  {{D'Orazio}}, \citenamefont {{Haiman}},\ and\ \citenamefont
  {{MacFadyen}}}]{2013MNRAS.436.2997D}%
  \BibitemOpen
  \bibfield  {author} {\bibinfo {author} {\bibfnamefont {D.~J.}\ \bibnamefont
  {{D'Orazio}}}, \bibinfo {author} {\bibfnamefont {Z.}~\bibnamefont
  {{Haiman}}}, \ and\ \bibinfo {author} {\bibfnamefont {A.}~\bibnamefont
  {{MacFadyen}}},\ }\href {\doibase 10.1093/mnras/stt1787} {\bibfield
  {journal} {\bibinfo  {journal} {Mon. Not. Roy. Astron. Soc.}\ }\textbf
  {\bibinfo {volume} {436}},\ \bibinfo {pages} {2997} (\bibinfo {year}
  {2013})},\ \Eprint {http://arxiv.org/abs/1210.0536} {arXiv:1210.0536
  [astro-ph.GA]} \BibitemShut {NoStop}%
\bibitem [{\citenamefont {{Canovas}}\ \emph {et~al.}(2017)\citenamefont
  {{Canovas}}, \citenamefont {{Hardy}}, \citenamefont {{Zurlo}}, \citenamefont
  {{Wahhaj}}, \citenamefont {{Schreiber}}, \citenamefont {{Vigan}},
  \citenamefont {{Villaver}}, \citenamefont {{Olofsson}}, \citenamefont
  {{Meeus}}, \citenamefont {{M{\'e}nard}}, \citenamefont {{Caceres}},
  \citenamefont {{Cieza}},\ and\ \citenamefont
  {{Garufi}}}]{2017A&A...598A..43C}%
  \BibitemOpen
  \bibfield  {author} {\bibinfo {author} {\bibfnamefont {H.}~\bibnamefont
  {{Canovas}}}, \bibinfo {author} {\bibfnamefont {A.}~\bibnamefont {{Hardy}}},
  \bibinfo {author} {\bibfnamefont {A.}~\bibnamefont {{Zurlo}}}, \bibinfo
  {author} {\bibfnamefont {Z.}~\bibnamefont {{Wahhaj}}}, \bibinfo {author}
  {\bibfnamefont {M.~R.}\ \bibnamefont {{Schreiber}}}, \bibinfo {author}
  {\bibfnamefont {A.}~\bibnamefont {{Vigan}}}, \bibinfo {author} {\bibfnamefont
  {E.}~\bibnamefont {{Villaver}}}, \bibinfo {author} {\bibfnamefont
  {J.}~\bibnamefont {{Olofsson}}}, \bibinfo {author} {\bibfnamefont
  {G.}~\bibnamefont {{Meeus}}}, \bibinfo {author} {\bibfnamefont
  {F.}~\bibnamefont {{M{\'e}nard}}}, \bibinfo {author} {\bibfnamefont
  {C.}~\bibnamefont {{Caceres}}}, \bibinfo {author} {\bibfnamefont {L.~A.}\
  \bibnamefont {{Cieza}}}, \ and\ \bibinfo {author} {\bibfnamefont
  {A.}~\bibnamefont {{Garufi}}},\ }\href {\doibase 10.1051/0004-6361/201629145}
  {\bibfield  {journal} {\bibinfo  {journal} {Astronomy \& Astrophysics}\
  }\textbf {\bibinfo {volume} {598}},\ \bibinfo {eid} {A43} (\bibinfo {year}
  {2017})},\ \Eprint {http://arxiv.org/abs/1606.07087} {arXiv:1606.07087
  [astro-ph.SR]} \BibitemShut {NoStop}%
\bibitem [{\citenamefont {Alonso-Monsalve}\ and\ \citenamefont
  {Kaiser}(2023{\natexlab{b}})}]{Alonso-Monsalve:2023jfq}%
  \BibitemOpen
  \bibfield  {author} {\bibinfo {author} {\bibfnamefont {E.}~\bibnamefont
  {Alonso-Monsalve}}\ and\ \bibinfo {author} {\bibfnamefont {D.~I.}\
  \bibnamefont {Kaiser}},\ }\href@noop {} {\  (\bibinfo {year}
  {2023}{\natexlab{b}})},\ \Eprint {http://arxiv.org/abs/2309.15385}
  {arXiv:2309.15385 [hep-ph]} \BibitemShut {NoStop}%
\bibitem [{\citenamefont {{Kaw}}\ and\ \citenamefont
  {{Dawson}}(1970)}]{KawDawson1970}%
  \BibitemOpen
  \bibfield  {author} {\bibinfo {author} {\bibfnamefont {P.}~\bibnamefont
  {{Kaw}}}\ and\ \bibinfo {author} {\bibfnamefont {J.}~\bibnamefont
  {{Dawson}}},\ }\href {\doibase 10.1063/1.1692942} {\bibfield  {journal}
  {\bibinfo  {journal} {Physics of Fluids}\ }\textbf {\bibinfo {volume} {13}},\
  \bibinfo {pages} {472} (\bibinfo {year} {1970})}\BibitemShut {NoStop}%
\bibitem [{\citenamefont {Regge}\ and\ \citenamefont
  {Wheeler}(1957)}]{Regge:1957td}%
  \BibitemOpen
  \bibfield  {author} {\bibinfo {author} {\bibfnamefont {T.}~\bibnamefont
  {Regge}}\ and\ \bibinfo {author} {\bibfnamefont {J.~A.}\ \bibnamefont
  {Wheeler}},\ }\href {\doibase 10.1103/PhysRev.108.1063} {\bibfield  {journal}
  {\bibinfo  {journal} {Phys. Rev.}\ }\textbf {\bibinfo {volume} {108}},\
  \bibinfo {pages} {1063} (\bibinfo {year} {1957})}\BibitemShut {NoStop}%
\bibitem [{\citenamefont {Zerilli}(1970{\natexlab{a}})}]{Zerilli:1970wzz}%
  \BibitemOpen
  \bibfield  {author} {\bibinfo {author} {\bibfnamefont {F.~J.}\ \bibnamefont
  {Zerilli}},\ }\href {\doibase 10.1103/PhysRevD.2.2141} {\bibfield  {journal}
  {\bibinfo  {journal} {Phys. Rev. D}\ }\textbf {\bibinfo {volume} {2}},\
  \bibinfo {pages} {2141} (\bibinfo {year} {1970}{\natexlab{a}})}\BibitemShut
  {NoStop}%
\bibitem [{\citenamefont {Zerilli}(1970{\natexlab{b}})}]{Zerilli:1970se}%
  \BibitemOpen
  \bibfield  {author} {\bibinfo {author} {\bibfnamefont {F.~J.}\ \bibnamefont
  {Zerilli}},\ }\href {\doibase 10.1103/PhysRevLett.24.737} {\bibfield
  {journal} {\bibinfo  {journal} {Phys. Rev. Lett.}\ }\textbf {\bibinfo
  {volume} {24}},\ \bibinfo {pages} {737} (\bibinfo {year}
  {1970}{\natexlab{b}})}\BibitemShut {NoStop}%
\bibitem [{\citenamefont {Zerilli}(1974)}]{Zerilli:1974ai}%
  \BibitemOpen
  \bibfield  {author} {\bibinfo {author} {\bibfnamefont {F.~J.}\ \bibnamefont
  {Zerilli}},\ }\href {\doibase 10.1103/PhysRevD.9.860} {\bibfield  {journal}
  {\bibinfo  {journal} {Phys. Rev. D}\ }\textbf {\bibinfo {volume} {9}},\
  \bibinfo {pages} {860} (\bibinfo {year} {1974})}\BibitemShut {NoStop}%
\bibitem [{\citenamefont {Pani}\ \emph {et~al.}(2013)\citenamefont {Pani},
  \citenamefont {Berti},\ and\ \citenamefont {Gualtieri}}]{Pani:2013wsa}%
  \BibitemOpen
  \bibfield  {author} {\bibinfo {author} {\bibfnamefont {P.}~\bibnamefont
  {Pani}}, \bibinfo {author} {\bibfnamefont {E.}~\bibnamefont {Berti}}, \ and\
  \bibinfo {author} {\bibfnamefont {L.}~\bibnamefont {Gualtieri}},\ }\href
  {\doibase 10.1103/PhysRevD.88.064048} {\bibfield  {journal} {\bibinfo
  {journal} {Phys. Rev. D}\ }\textbf {\bibinfo {volume} {88}},\ \bibinfo
  {pages} {064048} (\bibinfo {year} {2013})},\ \Eprint
  {http://arxiv.org/abs/1307.7315} {arXiv:1307.7315 [gr-qc]} \BibitemShut
  {NoStop}%
\bibitem [{\citenamefont {Rosa}\ and\ \citenamefont
  {Dolan}(2012)}]{Rosa:2011my}%
  \BibitemOpen
  \bibfield  {author} {\bibinfo {author} {\bibfnamefont {J.~G.}\ \bibnamefont
  {Rosa}}\ and\ \bibinfo {author} {\bibfnamefont {S.~R.}\ \bibnamefont
  {Dolan}},\ }\href {\doibase 10.1103/PhysRevD.85.044043} {\bibfield  {journal}
  {\bibinfo  {journal} {Phys. Rev. D}\ }\textbf {\bibinfo {volume} {85}},\
  \bibinfo {pages} {044043} (\bibinfo {year} {2012})},\ \Eprint
  {http://arxiv.org/abs/1110.4494} {arXiv:1110.4494 [hep-th]} \BibitemShut
  {NoStop}%
\bibitem [{\citenamefont {Baryakhtar}\ \emph {et~al.}(2017)\citenamefont
  {Baryakhtar}, \citenamefont {Lasenby},\ and\ \citenamefont
  {Teo}}]{Baryakhtar:2017ngi}%
  \BibitemOpen
  \bibfield  {author} {\bibinfo {author} {\bibfnamefont {M.}~\bibnamefont
  {Baryakhtar}}, \bibinfo {author} {\bibfnamefont {R.}~\bibnamefont {Lasenby}},
  \ and\ \bibinfo {author} {\bibfnamefont {M.}~\bibnamefont {Teo}},\ }\href
  {\doibase 10.1103/PhysRevD.96.035019} {\bibfield  {journal} {\bibinfo
  {journal} {Phys. Rev. D}\ }\textbf {\bibinfo {volume} {96}},\ \bibinfo
  {pages} {035019} (\bibinfo {year} {2017})},\ \Eprint
  {http://arxiv.org/abs/1704.05081} {arXiv:1704.05081 [hep-ph]} \BibitemShut
  {NoStop}%
\bibitem [{\citenamefont {Moncrief}(1974{\natexlab{a}})}]{Moncrief:1974ng}%
  \BibitemOpen
  \bibfield  {author} {\bibinfo {author} {\bibfnamefont {V.}~\bibnamefont
  {Moncrief}},\ }\href {\doibase 10.1103/PhysRevD.10.1057} {\bibfield
  {journal} {\bibinfo  {journal} {Phys. Rev. D}\ }\textbf {\bibinfo {volume}
  {10}},\ \bibinfo {pages} {1057} (\bibinfo {year}
  {1974}{\natexlab{a}})}\BibitemShut {NoStop}%
\bibitem [{\citenamefont {Moncrief}(1974{\natexlab{b}})}]{PhysRevD.9.2707}%
  \BibitemOpen
  \bibfield  {author} {\bibinfo {author} {\bibfnamefont {V.}~\bibnamefont
  {Moncrief}},\ }\href {\doibase 10.1103/PhysRevD.9.2707} {\bibfield  {journal}
  {\bibinfo  {journal} {Phys. Rev. D}\ }\textbf {\bibinfo {volume} {9}},\
  \bibinfo {pages} {2707} (\bibinfo {year} {1974}{\natexlab{b}})}\BibitemShut
  {NoStop}%
\bibitem [{\citenamefont {Moncrief}(1975)}]{Moncrief:1975sb}%
  \BibitemOpen
  \bibfield  {author} {\bibinfo {author} {\bibfnamefont {V.}~\bibnamefont
  {Moncrief}},\ }\href {\doibase 10.1103/PhysRevD.12.1526} {\bibfield
  {journal} {\bibinfo  {journal} {Phys. Rev. D}\ }\textbf {\bibinfo {volume}
  {12}},\ \bibinfo {pages} {1526} (\bibinfo {year} {1975})}\BibitemShut
  {NoStop}%
\bibitem [{\citenamefont {Zenginoglu}\ and\ \citenamefont
  {Khanna}(2011)}]{Zenginoglu:2011zz}%
  \BibitemOpen
  \bibfield  {author} {\bibinfo {author} {\bibfnamefont {A.}~\bibnamefont
  {Zenginoglu}}\ and\ \bibinfo {author} {\bibfnamefont {G.}~\bibnamefont
  {Khanna}},\ }\href {\doibase 10.1103/PhysRevX.1.021017} {\bibfield  {journal}
  {\bibinfo  {journal} {Phys. Rev. X}\ }\textbf {\bibinfo {volume} {1}},\
  \bibinfo {pages} {021017} (\bibinfo {year} {2011})},\ \Eprint
  {http://arxiv.org/abs/1108.1816} {arXiv:1108.1816 [gr-qc]} \BibitemShut
  {NoStop}%
\bibitem [{\citenamefont {Krivan}\ \emph {et~al.}(1997)\citenamefont {Krivan},
  \citenamefont {Laguna}, \citenamefont {Papadopoulos},\ and\ \citenamefont
  {Andersson}}]{Krivan_1997}%
  \BibitemOpen
  \bibfield  {author} {\bibinfo {author} {\bibfnamefont {W.}~\bibnamefont
  {Krivan}}, \bibinfo {author} {\bibfnamefont {P.}~\bibnamefont {Laguna}},
  \bibinfo {author} {\bibfnamefont {P.}~\bibnamefont {Papadopoulos}}, \ and\
  \bibinfo {author} {\bibfnamefont {N.}~\bibnamefont {Andersson}},\ }\href
  {\doibase 10.1103/physrevd.56.3395} {\bibfield  {journal} {\bibinfo
  {journal} {Physical Review D}\ }\textbf {\bibinfo {volume} {56}},\ \bibinfo
  {pages} {3395–3404} (\bibinfo {year} {1997})}\BibitemShut {NoStop}%
\bibitem [{\citenamefont {Pazos-Ávalos}\ and\ \citenamefont
  {Lousto}(2005)}]{Pazos_valos_2005}%
  \BibitemOpen
  \bibfield  {author} {\bibinfo {author} {\bibfnamefont {E.}~\bibnamefont
  {Pazos-Ávalos}}\ and\ \bibinfo {author} {\bibfnamefont {C.~O.}\ \bibnamefont
  {Lousto}},\ }\href {\doibase 10.1103/physrevd.72.084022} {\bibfield
  {journal} {\bibinfo  {journal} {Physical Review D}\ }\textbf {\bibinfo
  {volume} {72}} (\bibinfo {year} {2005}),\
  10.1103/physrevd.72.084022}\BibitemShut {NoStop}%
\bibitem [{\citenamefont {Zenginoglu}\ \emph {et~al.}(2014)\citenamefont
  {Zenginoglu}, \citenamefont {Khanna},\ and\ \citenamefont
  {Burko}}]{Zenginoglu:2012us}%
  \BibitemOpen
  \bibfield  {author} {\bibinfo {author} {\bibfnamefont {A.}~\bibnamefont
  {Zenginoglu}}, \bibinfo {author} {\bibfnamefont {G.}~\bibnamefont {Khanna}},
  \ and\ \bibinfo {author} {\bibfnamefont {L.~M.}\ \bibnamefont {Burko}},\
  }\href {\doibase 10.1007/s10714-014-1672-8} {\bibfield  {journal} {\bibinfo
  {journal} {Gen. Rel. Grav.}\ }\textbf {\bibinfo {volume} {46}},\ \bibinfo
  {pages} {1672} (\bibinfo {year} {2014})},\ \Eprint
  {http://arxiv.org/abs/1208.5839} {arXiv:1208.5839 [gr-qc]} \BibitemShut
  {NoStop}%
%%CITATION = ARXIV:1208.5839;%%
\bibitem [{\citenamefont {Cardoso}\ \emph
  {et~al.}(2021{\natexlab{b}})\citenamefont {Cardoso}, \citenamefont {Duque},\
  and\ \citenamefont {Khanna}}]{Cardoso:2021vjq}%
  \BibitemOpen
  \bibfield  {author} {\bibinfo {author} {\bibfnamefont {V.}~\bibnamefont
  {Cardoso}}, \bibinfo {author} {\bibfnamefont {F.}~\bibnamefont {Duque}}, \
  and\ \bibinfo {author} {\bibfnamefont {G.}~\bibnamefont {Khanna}},\ }\href
  {\doibase 10.1103/PhysRevD.103.L081501} {\bibfield  {journal} {\bibinfo
  {journal} {Phys. Rev. D}\ }\textbf {\bibinfo {volume} {103}},\ \bibinfo
  {pages} {L081501} (\bibinfo {year} {2021}{\natexlab{b}})},\ \Eprint
  {http://arxiv.org/abs/2101.01186} {arXiv:2101.01186 [gr-qc]} \BibitemShut
  {NoStop}%
\bibitem [{\citenamefont {Lingetti}\ \emph {et~al.}(2022)\citenamefont
  {Lingetti}, \citenamefont {Cannizzaro},\ and\ \citenamefont
  {Pani}}]{Lingetti:2022psy}%
  \BibitemOpen
  \bibfield  {author} {\bibinfo {author} {\bibfnamefont {G.}~\bibnamefont
  {Lingetti}}, \bibinfo {author} {\bibfnamefont {E.}~\bibnamefont
  {Cannizzaro}}, \ and\ \bibinfo {author} {\bibfnamefont {P.}~\bibnamefont
  {Pani}},\ }\href {\doibase 10.1103/PhysRevD.106.024007} {\bibfield  {journal}
  {\bibinfo  {journal} {Phys. Rev. D}\ }\textbf {\bibinfo {volume} {106}},\
  \bibinfo {pages} {024007} (\bibinfo {year} {2022})},\ \Eprint
  {http://arxiv.org/abs/2204.09335} {arXiv:2204.09335 [gr-qc]} \BibitemShut
  {NoStop}%
\bibitem [{\citenamefont {Dima}\ and\ \citenamefont
  {Barausse}(2020)}]{Dima:2020rzg}%
  \BibitemOpen
  \bibfield  {author} {\bibinfo {author} {\bibfnamefont {A.}~\bibnamefont
  {Dima}}\ and\ \bibinfo {author} {\bibfnamefont {E.}~\bibnamefont
  {Barausse}},\ }\href {\doibase 10.1088/1361-6382/ab9ce0} {\bibfield
  {journal} {\bibinfo  {journal} {Class. Quant. Grav.}\ }\textbf {\bibinfo
  {volume} {37}},\ \bibinfo {pages} {175006} (\bibinfo {year} {2020})},\
  \Eprint {http://arxiv.org/abs/2001.11484} {arXiv:2001.11484 [gr-qc]}
  \BibitemShut {NoStop}%
\bibitem [{\citenamefont {Cardoso}\ \emph
  {et~al.}(2016{\natexlab{b}})\citenamefont {Cardoso}, \citenamefont
  {Franzin},\ and\ \citenamefont {Pani}}]{Cardoso:2016rao}%
  \BibitemOpen
  \bibfield  {author} {\bibinfo {author} {\bibfnamefont {V.}~\bibnamefont
  {Cardoso}}, \bibinfo {author} {\bibfnamefont {E.}~\bibnamefont {Franzin}}, \
  and\ \bibinfo {author} {\bibfnamefont {P.}~\bibnamefont {Pani}},\ }\href
  {\doibase 10.1103/PhysRevLett.117.089902, 10.1103/PhysRevLett.116.171101}
  {\bibfield  {journal} {\bibinfo  {journal} {Phys. Rev. Lett.}\ }\textbf
  {\bibinfo {volume} {116}},\ \bibinfo {pages} {171101} (\bibinfo {year}
  {2016}{\natexlab{b}})},\ \bibinfo {note} {[Erratum: Phys. Rev.
  Lett.117,no.8,089902(2016)]},\ \Eprint {http://arxiv.org/abs/1602.07309}
  {arXiv:1602.07309 [gr-qc]} \BibitemShut {NoStop}%
%%CITATION = ARXIV:1602.07309;%%
\bibitem [{\citenamefont {Cardoso}\ \emph
  {et~al.}(2016{\natexlab{c}})\citenamefont {Cardoso}, \citenamefont {Hopper},
  \citenamefont {Macedo}, \citenamefont {Palenzuela},\ and\ \citenamefont
  {Pani}}]{Cardoso:2016oxy}%
  \BibitemOpen
  \bibfield  {author} {\bibinfo {author} {\bibfnamefont {V.}~\bibnamefont
  {Cardoso}}, \bibinfo {author} {\bibfnamefont {S.}~\bibnamefont {Hopper}},
  \bibinfo {author} {\bibfnamefont {C.~F.~B.}\ \bibnamefont {Macedo}}, \bibinfo
  {author} {\bibfnamefont {C.}~\bibnamefont {Palenzuela}}, \ and\ \bibinfo
  {author} {\bibfnamefont {P.}~\bibnamefont {Pani}},\ }\href {\doibase
  10.1103/PhysRevD.94.084031} {\bibfield  {journal} {\bibinfo  {journal} {Phys.
  Rev. D}\ }\textbf {\bibinfo {volume} {94}},\ \bibinfo {pages} {084031}
  (\bibinfo {year} {2016}{\natexlab{c}})},\ \Eprint
  {http://arxiv.org/abs/1608.08637} {arXiv:1608.08637 [gr-qc]} \BibitemShut
  {NoStop}%
\bibitem [{\citenamefont {Oshita}\ and\ \citenamefont
  {Afshordi}(2019)}]{Oshita:2018fqu}%
  \BibitemOpen
  \bibfield  {author} {\bibinfo {author} {\bibfnamefont {N.}~\bibnamefont
  {Oshita}}\ and\ \bibinfo {author} {\bibfnamefont {N.}~\bibnamefont
  {Afshordi}},\ }\href {\doibase 10.1103/PhysRevD.99.044002} {\bibfield
  {journal} {\bibinfo  {journal} {Phys. Rev. D}\ }\textbf {\bibinfo {volume}
  {99}},\ \bibinfo {pages} {044002} (\bibinfo {year} {2019})},\ \Eprint
  {http://arxiv.org/abs/1807.10287} {arXiv:1807.10287 [gr-qc]} \BibitemShut
  {NoStop}%
\bibitem [{\citenamefont {Wang}\ \emph {et~al.}(2020)\citenamefont {Wang},
  \citenamefont {Oshita},\ and\ \citenamefont {Afshordi}}]{Wang:2019rcf}%
  \BibitemOpen
  \bibfield  {author} {\bibinfo {author} {\bibfnamefont {Q.}~\bibnamefont
  {Wang}}, \bibinfo {author} {\bibfnamefont {N.}~\bibnamefont {Oshita}}, \ and\
  \bibinfo {author} {\bibfnamefont {N.}~\bibnamefont {Afshordi}},\ }\href
  {\doibase 10.1103/PhysRevD.101.024031} {\bibfield  {journal} {\bibinfo
  {journal} {Phys. Rev. D}\ }\textbf {\bibinfo {volume} {101}},\ \bibinfo
  {pages} {024031} (\bibinfo {year} {2020})},\ \Eprint
  {http://arxiv.org/abs/1905.00446} {arXiv:1905.00446 [gr-qc]} \BibitemShut
  {NoStop}%
\bibitem [{\citenamefont {Ferrari}\ and\ \citenamefont
  {Kokkotas}(2000)}]{Ferrari:2000sr}%
  \BibitemOpen
  \bibfield  {author} {\bibinfo {author} {\bibfnamefont {V.}~\bibnamefont
  {Ferrari}}\ and\ \bibinfo {author} {\bibfnamefont {K.~D.}\ \bibnamefont
  {Kokkotas}},\ }\href {\doibase 10.1103/PhysRevD.62.107504} {\bibfield
  {journal} {\bibinfo  {journal} {Phys. Rev. D}\ }\textbf {\bibinfo {volume}
  {62}},\ \bibinfo {pages} {107504} (\bibinfo {year} {2000})},\ \Eprint
  {http://arxiv.org/abs/gr-qc/0008057} {arXiv:gr-qc/0008057} \BibitemShut
  {NoStop}%
\bibitem [{\citenamefont {Pani}\ and\ \citenamefont
  {Ferrari}(2018)}]{Pani:2018flj}%
  \BibitemOpen
  \bibfield  {author} {\bibinfo {author} {\bibfnamefont {P.}~\bibnamefont
  {Pani}}\ and\ \bibinfo {author} {\bibfnamefont {V.}~\bibnamefont {Ferrari}},\
  }\href {\doibase 10.1088/1361-6382/aacb8f} {\bibfield  {journal} {\bibinfo
  {journal} {Class. Quant. Grav.}\ }\textbf {\bibinfo {volume} {35}},\ \bibinfo
  {pages} {15LT01} (\bibinfo {year} {2018})},\ \Eprint
  {http://arxiv.org/abs/1804.01444} {arXiv:1804.01444 [gr-qc]} \BibitemShut
  {NoStop}%
\bibitem [{\citenamefont {Buoninfante}\ and\ \citenamefont
  {Mazumdar}(2019)}]{Buoninfante:2019swn}%
  \BibitemOpen
  \bibfield  {author} {\bibinfo {author} {\bibfnamefont {L.}~\bibnamefont
  {Buoninfante}}\ and\ \bibinfo {author} {\bibfnamefont {A.}~\bibnamefont
  {Mazumdar}},\ }\href {\doibase 10.1103/PhysRevD.100.024031} {\bibfield
  {journal} {\bibinfo  {journal} {Phys. Rev. D}\ }\textbf {\bibinfo {volume}
  {100}},\ \bibinfo {pages} {024031} (\bibinfo {year} {2019})},\ \Eprint
  {http://arxiv.org/abs/1903.01542} {arXiv:1903.01542 [gr-qc]} \BibitemShut
  {NoStop}%
\bibitem [{\citenamefont {Buoninfante}\ \emph {et~al.}(2019)\citenamefont
  {Buoninfante}, \citenamefont {Mazumdar},\ and\ \citenamefont
  {Peng}}]{Buoninfante:2019teo}%
  \BibitemOpen
  \bibfield  {author} {\bibinfo {author} {\bibfnamefont {L.}~\bibnamefont
  {Buoninfante}}, \bibinfo {author} {\bibfnamefont {A.}~\bibnamefont
  {Mazumdar}}, \ and\ \bibinfo {author} {\bibfnamefont {J.}~\bibnamefont
  {Peng}},\ }\href {\doibase 10.1103/PhysRevD.100.104059} {\bibfield  {journal}
  {\bibinfo  {journal} {Phys. Rev. D}\ }\textbf {\bibinfo {volume} {100}},\
  \bibinfo {pages} {104059} (\bibinfo {year} {2019})},\ \Eprint
  {http://arxiv.org/abs/1906.03624} {arXiv:1906.03624 [gr-qc]} \BibitemShut
  {NoStop}%
\bibitem [{\citenamefont {Delhom}\ \emph {et~al.}(2019)\citenamefont {Delhom},
  \citenamefont {Macedo}, \citenamefont {Olmo},\ and\ \citenamefont
  {Crispino}}]{Delhom:2019btt}%
  \BibitemOpen
  \bibfield  {author} {\bibinfo {author} {\bibfnamefont {A.}~\bibnamefont
  {Delhom}}, \bibinfo {author} {\bibfnamefont {C.~F.~B.}\ \bibnamefont
  {Macedo}}, \bibinfo {author} {\bibfnamefont {G.~J.}\ \bibnamefont {Olmo}}, \
  and\ \bibinfo {author} {\bibfnamefont {L.~C.~B.}\ \bibnamefont {Crispino}},\
  }\href {\doibase 10.1103/PhysRevD.100.024016} {\bibfield  {journal} {\bibinfo
   {journal} {Phys. Rev. D}\ }\textbf {\bibinfo {volume} {100}},\ \bibinfo
  {pages} {024016} (\bibinfo {year} {2019})},\ \Eprint
  {http://arxiv.org/abs/1906.06411} {arXiv:1906.06411 [gr-qc]} \BibitemShut
  {NoStop}%
\bibitem [{\citenamefont {Zhang}\ and\ \citenamefont
  {Zhou}(2018)}]{Zhang:2017jze}%
  \BibitemOpen
  \bibfield  {author} {\bibinfo {author} {\bibfnamefont {J.}~\bibnamefont
  {Zhang}}\ and\ \bibinfo {author} {\bibfnamefont {S.-Y.}\ \bibnamefont
  {Zhou}},\ }\href {\doibase 10.1103/PhysRevD.97.081501} {\bibfield  {journal}
  {\bibinfo  {journal} {Phys. Rev. D}\ }\textbf {\bibinfo {volume} {97}},\
  \bibinfo {pages} {081501} (\bibinfo {year} {2018})},\ \Eprint
  {http://arxiv.org/abs/1709.07503} {arXiv:1709.07503 [gr-qc]} \BibitemShut
  {NoStop}%
\bibitem [{\citenamefont {Cardoso}\ and\ \citenamefont
  {Pani}(2017)}]{Cardoso:2017cqb}%
  \BibitemOpen
  \bibfield  {author} {\bibinfo {author} {\bibfnamefont {V.}~\bibnamefont
  {Cardoso}}\ and\ \bibinfo {author} {\bibfnamefont {P.}~\bibnamefont {Pani}},\
  }\href {\doibase 10.1038/s41550-017-0225-y} {\bibfield  {journal} {\bibinfo
  {journal} {Nature Astron.}\ }\textbf {\bibinfo {volume} {1}},\ \bibinfo
  {pages} {586} (\bibinfo {year} {2017})},\ \Eprint
  {http://arxiv.org/abs/1709.01525} {arXiv:1709.01525 [gr-qc]} \BibitemShut
  {NoStop}%
\bibitem [{\citenamefont {Starobinskil}\ and\ \citenamefont
  {Churilov}(1974)}]{Starobinskil:1974nkd}%
  \BibitemOpen
  \bibfield  {author} {\bibinfo {author} {\bibfnamefont {A.~A.}\ \bibnamefont
  {Starobinskil}}\ and\ \bibinfo {author} {\bibfnamefont {S.~M.}\ \bibnamefont
  {Churilov}},\ }\href@noop {} {\bibfield  {journal} {\bibinfo  {journal} {Sov.
  Phys. JETP}\ }\textbf {\bibinfo {volume} {65}},\ \bibinfo {pages} {1}
  (\bibinfo {year} {1974})}\BibitemShut {NoStop}%
\bibitem [{\citenamefont {Brito}\ \emph {et~al.}(2014)\citenamefont {Brito},
  \citenamefont {Cardoso},\ and\ \citenamefont {Pani}}]{Brito:2014nja}%
  \BibitemOpen
  \bibfield  {author} {\bibinfo {author} {\bibfnamefont {R.}~\bibnamefont
  {Brito}}, \bibinfo {author} {\bibfnamefont {V.}~\bibnamefont {Cardoso}}, \
  and\ \bibinfo {author} {\bibfnamefont {P.}~\bibnamefont {Pani}},\ }\href
  {\doibase 10.1103/PhysRevD.89.104045} {\bibfield  {journal} {\bibinfo
  {journal} {Phys. Rev. D}\ }\textbf {\bibinfo {volume} {89}},\ \bibinfo
  {pages} {104045} (\bibinfo {year} {2014})},\ \Eprint
  {http://arxiv.org/abs/1405.2098} {arXiv:1405.2098 [gr-qc]} \BibitemShut
  {NoStop}%
\bibitem [{\citenamefont {Pere\~niguez}(2023)}]{Pereniguez:2023wxf}%
  \BibitemOpen
  \bibfield  {author} {\bibinfo {author} {\bibfnamefont {D.}~\bibnamefont
  {Pere\~niguez}},\ }\href {\doibase 10.1103/PhysRevD.108.084046} {\bibfield
  {journal} {\bibinfo  {journal} {Phys. Rev. D}\ }\textbf {\bibinfo {volume}
  {108}},\ \bibinfo {pages} {084046} (\bibinfo {year} {2023})},\ \Eprint
  {http://arxiv.org/abs/2302.10942} {arXiv:2302.10942 [gr-qc]} \BibitemShut
  {NoStop}%
\bibitem [{\citenamefont {Dyson}\ and\ \citenamefont
  {Pere\~niguez}(2023)}]{Dyson:2023ujk}%
  \BibitemOpen
  \bibfield  {author} {\bibinfo {author} {\bibfnamefont {C.}~\bibnamefont
  {Dyson}}\ and\ \bibinfo {author} {\bibfnamefont {D.}~\bibnamefont
  {Pere\~niguez}},\ }\href {\doibase 10.1103/PhysRevD.108.084064} {\bibfield
  {journal} {\bibinfo  {journal} {Phys. Rev. D}\ }\textbf {\bibinfo {volume}
  {108}},\ \bibinfo {pages} {084064} (\bibinfo {year} {2023})},\ \Eprint
  {http://arxiv.org/abs/2306.15751} {arXiv:2306.15751 [gr-qc]} \BibitemShut
  {NoStop}%
\end{thebibliography}%
\clearpage
\renewcommand{\thesubsection}{{S.\arabic{subsection}}}
\setcounter{section}{0}
%%%%%%%%%%%%%%%%%%%%%%%%%%%%%%%%%%
\section*{Supplemental material}
%%%%%%%%%%%%%%%%%%%%%%%%%%%%%%%%%%
%%%%%%%%%%%%%%%%%%%%%%%%%%%%%%%%%
\subsection{Perturbation theory}\label{app:perturbationtheory}
%%%%%%%%%%%%%%%%%%%%%%%%%%%%%%%%%
We provide details of our perturbation scheme and outline the relevant steps to obtain the evolution equations \eqref{eq:wavelike-eqn}, within the non-relativistic fluid approximation. We find it convenient to perform computations in the frequency domain by assuming a sinusoidal time dependence, and switch back to the time domain once we obtain the master equations. Note that, as the system is stationary, it is possible to go back and forth between them simply through $\partial/\partial t \leftrightarrow -i \omega$. 

From the background solution \eqref{eq:RN_Sol}, we linearize the field equations \eqref{eq:einstein_maxwell} to first order perturbations. Opposed to vacuum RN, where one only needs to perturb the gravitational and EM field, we need to consider perturbations of the fluid quantities as well. Given the large hierarchy between the mass of the electrons and ions, we ignore perturbations on the latter and treat them as a stationary, neutralizing background. Our perturbation scheme is thus given by $X=X^{(0)}+\epsilon\,\delta\! X+\mathcal{O}(\epsilon^2)$, where $X=g_{\alpha\beta}, A_{\alpha}, v_{\alpha}, n_{\rm e}, P_{\rm t}$, the superscript $(0)$ denotes background quantities and $\epsilon$ is a bookkeeping parameter. Due to the spherical symmetry of the background solution, the angular dependence from any linear perturbation can be separated through a multipolar expansion. 

The gravitational perturbations are expanded in a basis of tensor spherical harmonics, which fall into an axial or polar category, depending on their behaviour under parity \cite{Regge:1957td,Zerilli:1970se,Zerilli:1970wzz}. We thus write
\begin{equation}
    \delta g_{\alpha \beta}= \delta g_{\alpha \beta}^{\rm axial}+\delta g_{\alpha \beta}^{\rm polar}\,,
\end{equation}
which in the Regge-Wheeler gauge can be written as
\begin{align}
\label{eq:Regge-Wheeler-gauge}
\renewcommand{\arraystretch}{1.25}
\setlength\arraycolsep{2pt}
\delta g_{\alpha \beta}\!=\!\sum_{\ell, m}\begin{pNiceArray}{cc|cc}[margin = 0.cm]
\Block[fill=blue!10,rounded-corners]{2-2}{}
H_0Y & H_1Y & \Block[fill=red!10,rounded-corners]{2-2}{}
        h_0 S_{\theta} & h_0 S_{\varphi} \\
H_1Y & H_2Y & h_1S_{\theta} & h_1 S_{\varphi} \\
\hline
\Block[fill=red!10,rounded-corners]{2-2}{}
* & * & \Block[fill=blue!10,rounded-corners]{2-2}{}
        r^2 KY & 0 \\
* & * & 0 & r^{2}\sin^{2}{\theta} KY \\
\end{pNiceArray}e^{-i \omega t}\,.
\end{align}
Here, the red and blue blocks correspond to the axial $(\delta g_{\alpha \beta}^{\rm axial})$ and polar part $(\delta g_{\alpha \beta}^{\rm polar})$, respectively, and to avoid cluttering, we omit all dependencies on $(t,r,\theta,\varphi)$ as well as the $(\ell,m)$ label on the radial functions in the axial ($h^{\ell m}_0$ and $h^{\ell m}_1$) and polar ($H^{\ell m}_0$, $H^{\ell m}_1$, $H^{\ell m}_2$ and $K^{\ell m}$) sector, the spherical harmonics $Y = Y^{\ell m}$ and the axial vector harmonics $(S_{\theta}, S_{\varphi})=(S^{\ell m}_{\theta}, S^{\ell m}_{\varphi})  = (-\partial_{\varphi}Y^{\ell m}/\sin{\theta},\sin{\theta}\partial_{\theta}Y^{\ell m})$. Finally, we define $\sum_{\ell,m} \equiv \sum_{\ell = 0}^{\infty}\sum_{m = -\ell}^{\ell}$. 

The EM field and the four-velocity can be expanded in a basis of vector spherical harmonics as
\begin{equation}
\label{eq:expansions}
\begin{aligned}
    \delta A_\alpha&=\frac{1}{r}\sum_{i=1}^4\sum_{\ell,m}c_iu_{i}^{\ell m}  Z_\alpha^{(i) \ell m} e^{-i \omega t}\,,\\
    \delta v_\alpha&=\frac{1}{r}\sum_{i=1}^4\sum_{\ell ,m}c_i v_{i}^{\ell m}  Z_\alpha^{(i)\ell m}e^{-i \omega t}\,,
\end{aligned}
\end{equation}
where $c_1=c_2=1$ and $c_3=c_4=1/\sqrt{\ell(\ell+1)}$ and 
$Z_\alpha^{\ell m}$ is defined in e.g.~\cite{Rosa:2011my}. 

Finally, the density and pressure are expanded with the spherical harmonics due to their scalar nature:
\begin{equation}
\begin{aligned}
\delta n_{\rm e} &= \sum_{\ell,m} n^{\ell m}_{\mathrm{e}} Y^{\ell m} e^{-i \omega t}\,,\\
\delta P_{\rm t} &= \sum_{\ell,m} P^{\ell m}_{\mathrm{t}} Y^{\ell m} e^{-i \omega t}\,.
\end{aligned}
\end{equation}
%

%%%%%%%%%%%%%%%%%%%%%%%%%%
\subsubsection{Axial sector}
%%%%%%%%%%%%%%%%%%%%%%%%%%
The evolution of the perturbations in the axial sector can be written as a set of two coupled wave equations. This procedure is aided by the existence of a simple relation between the axial components of the plasma velocity and the EM field, found by perturbing the angular components of the momentum equation~\eqref{eq:momentum_continuity}:\footnote{We remind the reader that $v_{i}$ are the coefficients in the expansion of the four-velocity of the electrons while $u_{i}$ are the coefficients from the EM field, see \eqref{eq:expansions}.} 
\begin{equation}
\label{eq:momentumcons}
v_{4}=-\frac{e}{m_{\rm e}}u_{4}\,.
\end{equation}
The above equation simply corresponds to the conservation of the generalized transverse momentum of a charged particle with impulse $\vec{P}$ in an EM field $\vec{A}$, i.e., $\partial_t (\vec{P}+ e \vec{A})=0$. Given that the axial sector is composed solely of transverse modes~\cite{Baryakhtar:2017ngi}, this very simple relation holds for our purposes. Moreover, by considering only axial quantities, the continuity equation~\eqref{eq:momentum_continuity} and the radial component of the momentum equation show that $n^{\ell m}_{\mathrm{e}}=P^{\ell m}_{\mathrm{t}}=0$, which is a consequence of variations in the fluid density and pressure being longitudinal---and thus strictly polar---degrees of freedom. 

Fluctuations of the fluid's stress-energy tensor affect the axial Einstein equations via terms proportional to $\omega_{\rm p} m_{\rm e}/e$ and $\omega_{\rm p} ^2 m_{\rm e}^2/e^2$. Given that in astrophysical systems of interest $\omega_{\rm p}=\mathcal{O}(1/M)$, the large charge-to mass ratio of the electron ($e/m_{\rm e} \approx 10^{22}$) drastically suppresses the impact of the fluid on the gravitational sector, which we consequently neglect. 

This leaves us with three degrees of freedom, $h_0, h_1$ and $u_{4}$, for which there are three coupled equations (two of which are gravitational and one that is EM). The first gravitational equation comes from the $t\!-\!\theta$ and $t\!-\!\varphi$ components of Einstein equation, while the second one can be derived from the $r\!-\!\theta$ and $r\!-\!\varphi$ components. The Einstein equations are not modified by plasma and one can solve the latter to obtain:
\begin{equation}
    \frac{\partial h_0}{\partial r}= 2 \frac{h_0}{r}-\frac{4 Q u_{4}}{\lambda r^2}+\frac{i\Big(-2+\lambda-r^2\omega^2/f\Big)}{r\omega}\Psi_{\rm RW}\,,
\end{equation}
where we defined the Regge-Wheeler ``master variable'' $\Psi_{\rm RW}= f h_1 /r$. We can then use this expression in the first gravitational equation, which reduces to a simple relation between $h_0$ and $\Psi_{\rm RW}$:
\begin{equation}
    h_0=\frac{f}{-i \omega}\frac{\partial (r\Psi_{\rm RW})}{\partial r}\,.
\end{equation}
As $h_0$ is now decoupled from the system, we are able to obtain two coupled equations for the ``master variables'' $\Psi_{\rm RW}$ and $u_4$ as
\begin{equation}
\label{eq:wavelikeRW}
\begin{aligned}
&\frac{\partial^{2}\Psi_{\rm RW}}{\partial r^{2}}-\frac{2(Q^2- M r)}{r^{3}f}\frac{\partial\Psi_{\rm RW}}{\partial r} - \frac{4iQ\omega}{r^{3}\lambda f}u_{4} \\&+ \left(\frac{\omega^{2}}{f^{2}} - \frac{r^2 \lambda -6 M r + 4 Q^2}{r^{4}f}  \right)\Psi_{\rm RW}  = 0\,,\\[6pt]
&\frac{\partial^{2}u_{4}}{\partial r^{2}} - \frac{2(Q^{2}-Mr)}{r^{3}f}\frac{\partial u_4}{\partial r} +\frac{iQ\lambda(\lambda-2)}{r^{3}\omega f}\Psi_{\rm RW} \\& + \frac{1}{r^{4}f^{2}}\left(r^{4}\omega^{2} - (4 Q^2+r^2\lambda+\omega_{\rm p}^2 r^4)f\right)u_4 = 0\,.
\end{aligned}
\end{equation}
The full system has thus reduced to a set of coupled ordinary differential equations for $\Psi_{\rm RW}$ and $u_{4}$. To study the dynamics emerging from these equations, we transform them back to the time-domain. Furthermore, to acquire a wave-like form of the equations, we consider the Moncrief master variable $\Psi$ instead of the Regge-Wheeler one. They are related through \cite{Moncrief:1974ng,PhysRevD.9.2707,Moncrief:1975sb}: 
\begin{equation}
\label{eq:Moncrief}
\Psi= \frac{2 i \Psi_{\rm RW}}{\omega}\,.
\end{equation}
Using this substitution, we can rewrite \eqref{eq:wavelikeRW} as the wave equations reported in the main text \eqref{eq:wavelike-eqn}.
%%%%%%%%%%%%%%%%%%%%%%%%%%
\subsubsection{Polar sector}\label{app:polar}
%%%%%%%%%%%%%%%%%%%%%%%%%%
In contrast to the axial sector, fluid and EM perturbations are coupled in the polar sector. This is due the presence of longitudinal, electrostatic modes as well as gravitational fluctuations in the momentum equation~\eqref{eq:momentum_continuity}. Additionally, the polar density and pressure fluctuations $\delta n_{\rm e}^{\ell m}, \delta P_{\rm t}^{\ell m}$ do not vanish, while the perturbed velocity does not decouple trivially. There is thus no equivalent to Eq.~\eqref{eq:momentumcons}, which complicates the procedure.

In order to study the polar sector consistently, one should evolve the fluid variables along with the gravitational and EM fields and close the system with a suitable equation of state, as done in e.g.~\cite{Cardoso:2022whc}. We do not undertake that exercise here yet rather show that, similar to the axial sector, EM modes in the polar sector are dressed with an effective mass given by $\omega_{\rm p}$. Since the results in this work are centered around the presence of such a mass, there is no need to repeat the same detailed study in the polar sector as we did in the axial sector.

In the following, we briefly sketch the procedure to show the presence of an effective mass $\omega_{\rm p}$, in the polar sector. To obtain the EM master equation, one should consider the perturbation of the normalization of the four velocity, $\delta(v_\mu v^\mu)=0$ and solve for $v_1$ in terms of metric perturbations. This relation can then be substituted in the $t\!-\!r$ component of the Einstein equations, which gives $v_2$ in terms of metric perturbations. Finally, from the angular components of the momentum equation, one can obtain $v_3$ in terms of metric, EM and density perturbations. Next, consider the polar Maxwell's equations, and in particular the radial one---which we shall denote $\mathcal{M}_r$---and a combination of the angular ones, given by $\mathcal{M}_\theta+\mathcal{M}_\phi/\text{sin}^2\theta$. In analogy to previous work \cite{Zerilli:1970se, Pani:2018flj}, the EM master relation can be derived from these equations. Following Zerilli, we find it convenient to use perturbations of the field strength $\delta F_{\mu\nu}=\partial_\mu\,\delta A_\nu-\partial_\nu\,\delta A_\mu$, rather than the EM potential $\delta A^\mu$, as a dynamical variable. We choose a gauge such that $u_3=0$, while the other components of the potential are expressed as
\begin{equation}
\begin{aligned}
    u_1&= r\,\delta \Tilde{F}_{02}\,, \quad    u_2= r f\, \delta\Tilde{F}_{12}\,,\\
    \frac{\partial u_1}{\partial r}&= r\,\delta \Tilde{F}_{01}+\delta \Tilde{F}_{02}-i r \omega\,\delta \Tilde{F}_{02}\,,
\end{aligned}
\end{equation}
where $\delta \Tilde{F}_{\mu\nu}$ is the angle-independent part of $\delta F_{\mu\nu}$.
Using this relation, one can solve $\mathcal{M}_r$ and $\mathcal{M}_\theta+\mathcal{M}_\phi/\text{sin}^2\theta$ to obtain $\delta \Tilde{F}_{01}, \delta \Tilde{F}_{02}$, respectively. These solutions can then be used in the homogeneous Maxwell equation,
\begin{equation}
    \delta\Tilde{F}_{01}=\frac{ \partial\,\delta \Tilde{F}_{02}}{\partial r}+i \omega\,\delta \Tilde{F}_{12}\,,
\end{equation}
to obtain a second order differential equation for $\delta \Tilde{F}_{12}$. At large radii and neglecting the coupling with metric and density perturbations induced by the fluid, one finds the standard dispersion relation of a transverse EM mode in a plasma dressed by an effective mass given by the plasma frequency $\omega^2= k^2+\omega_{\rm p}^2$.

%%%%%%%%%%%%%%%%%%%%%%%%%%
\subsection{Relativistic regime}\label{app:rel-regime}
%%%%%%%%%%%%%%%%%%%%%%%%%%
We can relax the assumption of a non-relativistic fluid and show that the full momentum equation gives rise to a relativistic transparency effect induced by the BH charge. Analogous to the non-relativistic case, we solve the background momentum equation to find
\begin{equation}
    P_{\rm t}=-\frac{n_{\rm e}\left(e Q r\sqrt{f}+(Q^2-Mr)m_{\rm e}\right)}{2fr^{2}}\,,
\end{equation}
while we solve the perturbed axial momentum equation to obtain a relationship between the perturbed four-velocity and the EM axial mode:
\begin{equation}
    v_4=\frac{2 e r^2 f}{e Q r\sqrt{f}-\left(Q^2-3Mr+2r^{2}\right)m_{\rm e}}u_4 \, .
\end{equation}
In the limit of large $r$, i.e., in the non-relativistic regime, the above relation correctly reduces to Eq.~\eqref{eq:momentumcons} and thus one recovers $ e n^{(0)}_{\rm e} v_4=-\omega_{\rm p}^2 u_4$. In the strong field regime instead, where $e\gg m_{\rm e}$, the current in the Maxwell equations yields
\begin{equation}
    e n^{(0)}_{\rm e}v_4 \approx \frac{2 e n^{(0)}_{\rm e} r\sqrt{f}}{Q} u_4\,.
\end{equation}
In other words, we find an effective plasma frequency
\begin{equation}
\label{eq:effectiveplasmafreq}
\omega_{\rm p}^2= \frac{2 e n^{(0)}_{\rm e} \sqrt{f}}{A_0}\,,
\end{equation}
where we define the EM potential of the BH $A_{0}=Q/r$.

Compare this to a well-known result from plasma physics where in the presence of a strong EM potential $\vec{A}$, electrons are unable to oscillate due to their large relativistic mass. In this scenario, the {\it relativistic} plasma frequency reads~\cite{KawDawson1970}
\begin{equation}
\label{eq:plasmafreqKaw}
    \omega_{\rm p}^2= \frac{\tilde{n}_{\rm e} e^2}{\tilde{m}_{\rm e} \gamma_{\rm e}}\,,
\end{equation}
where $\gamma_{\rm e}$ is the Lorentz factor of the electrons under the influence of an EM field, while $\tilde{m}_{\rm e}$ and $\tilde{n}_{\rm e}$ are the rest mass and rest mass density, respectively. In our model, the Lorentz factor can be estimated by equating the (relativistic) centrifugal force of the circular orbits of the electrons to the Coulomb attraction between the electrons and the charged BH. This yields $\gamma_{\rm e}\approx \sqrt{1+ A_0 e/ \tilde{m}_{\rm e}}$. 

Crucially, if one uses $e A_0/\tilde{m}_{\rm e}\gg 1$ and $\tilde{n}_{\rm e}=n_{\rm e}/\gamma_{\rm e}$, Eqs.~\eqref{eq:plasmafreqKaw} and \eqref{eq:effectiveplasmafreq} coincide (modulo a redshift factor). Thus we find a relativistic transparency effect that suppresses the plasma frequency in the vicinity of the BH. This is the first time, to the best of our knowledge, that a consistent model of plasmas around charged BHs leads to this effect. The analysis done in the main text can therefore be trivially generalized to the relativistic regime by considering this relativistically-corrected effective mass.

%%%%%%%%%%%%%%%%%%%%%%%%%%
\subsection{Plasma features}\label{app:plasma}
%%%%%%%%%%%%%%%%%%%%%%%%%%
The plasma profile used throughout this work is defined in \eqref{eq:plasma-profile2}. Here, we aim to provide more details on its main features. Similar models were used in \cite{Cannizzaro:2020uap,Cannizzaro:2021zbp,Cannizzaro:2023ltu,Spieksma:2023vwl}, where further details can be found. 

In the uncharged case $\overline{Q} = 0$, the gravitational and EM sector decouple and thus the plasma can only affect the latter. Initializing the system with $\mathrm{ID}_{2}$, while placing the plasma away from the BH, we vary the plasma frequency to understand its impact on the EM sector. 
\begin{figure}
    \centering
    \includegraphics[width = 0.95\linewidth]{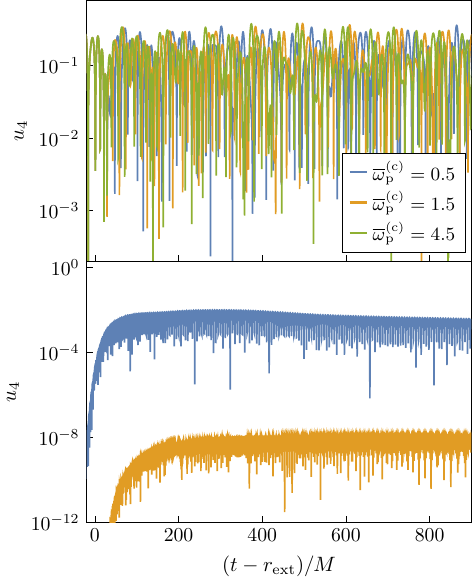}
    \caption{We show the EM sector $u_{4}$ with a plasma at $r_{\rm cut} = 50M$, while initializing in both sectors $\mathrm{ID}_{2}$ and varying the amplitude of the plasma barrier $\omega^{\rm (c)}_{\rm p}$. In the top panel, $u_{4}$ is extracted at $r_{\rm ext} = 30M$ and we see trapped waves in the cavity. In the bottom panel, the EM component is extracted at $r_{\rm ext} = 300M$. As expected, while increasing the plasma barrier, less radiation is able to travel through. In fact, we do not show $\overline{\omega}^{\rm (c)}_{\rm p} = 4.5$, as it reaches the noise level.}
    \label{fig:unchargedplasma}
\end{figure}
In Fig.~\ref{fig:unchargedplasma}, we show said system both at small ({\it top panel}) and large radii ({\it bottom panel}). In both cases, the formation of long-lived modes can be seen, yet these have a different origin. At small radii, there are modes that do not decay in time and have frequency {\it smaller} than the plasma frequency. These are EM waves that are trapped between the gravitational potential and the plasma barrier \cite{Lingetti:2022psy,Dima:2020rzg}. At large radii, we find modes with a frequency {\it higher} than the plasma frequency, which confirms these are travelling waves. They originate from the initial Gaussian, which has a tail with frequencies $\omega > \omega_{\rm p}^{\rm (c)}$. Indeed, by increasing the plasma frequency, these modes become less and less prominent, as shown in the bottom panel of Fig.~\ref{fig:unchargedplasma}. This proves how plasma acts as a high-pass filter, with a critical threshold given by the plasma frequency.
%%%%%%%%%%%%%%%%%%%%%%%%%%%%%%
\subsection{Numerical convergence}\label{app:convergence}
\begin{figure}
    \centering
    \includegraphics[width = 0.95\linewidth]{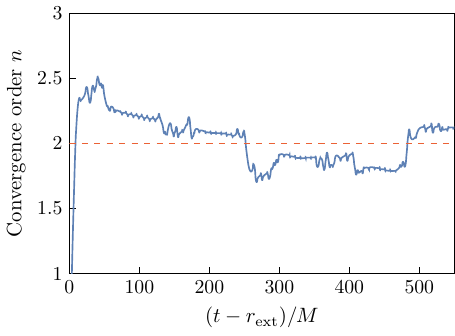}
    \caption{Convergence order as measured with $\mathrm{d}x/M =  \{0.1, 0.05, 0.025\}$. The result is consistent with the expected second-order convergence rate. We did the convergence test on a run with both $Q \neq 0$ and $\omega_{\rm p} \neq 0$.}
    \label{fig:convergence}
\end{figure}

%%%%%%%%%%%%%%%%%%%%%%%%%%%%%%
We check convergence of our time-domain approach in Fig.~\ref{fig:convergence}. We show the Cauchy convergence order $n = \log_{2}(||\Psi_{2h} - \Psi_{h}||_{2}\,/\, ||\Psi_{h}-\Psi_{h/2}||_{2})$, where we apply a moving average to smooth out the curve. The shown convergence rate is consistent with the second-order finite differences scheme that we use. All the simulations shown in this work are with $\Psi_h$, which has $\mathrm{d}x = 0.05M$, while we pick $\mathrm{d}t = 0.5 \mathrm{d}x$ such that the Courant–Friedrichs–Lewy condition is always satisfied.
%%%%%%%%%%%%%%%%%%%%%%%%%%%%%%
\end{document}